\documentclass[10pt]{iopart}
\usepackage{graphicx}
\usepackage{color}
\usepackage[utf8]{inputenc}
\usepackage[T1]{fontenc}
\usepackage{lmodern}
\usepackage[right]{rotating}
\usepackage{longtable}
\usepackage{array}
\usepackage{multirow}
\usepackage{paralist}
\usepackage[margin=10pt,font=small,labelfont=bf,
labelsep=endash]{caption}
\usepackage{subcaption}
\usepackage[numbers,sort&compress]{natbib}
\newcommand{\diff}{\mathrm{d}}
\newcommand{\codename}[1]{\texttt{#1}}
\newcommand{\beq}[1]{\begin{equation} #1 \end{equation}}

\graphicspath{{./}}

\begin{document}
\title[]{Binary neutron star merger simulations with different initial orbital frequency and equation of state.}
\author{F Maione$^1$, R De Pietri$^1$, A Feo$^1$ and F L\"offler$^2$}
\address{$^1$ Parma University and INFN Parma, Parco Area delle Scienze 7/A, I-43124 Parma (PR), Italy}
\address{$^2$ Center for Computation \& Technology, Louisiana State University, Baton Rouge, LA 70803 USA}

\begin{abstract}
We present results from three-dimensional general relativistic
simulations of binary neutron star coalescences and mergers using public 
codes.  We considered equal mass models where the baryon mass of the two 
Neutron Stars (NS) 
is $1.4M_{\odot}$, described by four different equations of state (EOS) for the
cold nuclear matter (APR4, SLy, H4, and MS1; all parametrized as
piecewise polytropes). We started the simulations from four different initial 
interbinary distances ($40, 44.3, 50$, and $60$ km), including up to the 
last 16 orbits before merger. That allows to show the effects on the gravitational wave phase evolution,
radiated energy and angular momentum due to: the use of different EOSs, 
the orbital eccentricity present in the initial data and  the initial 
separation (in the simulation) between the two stars.

Our results show that eccentricity has a major role in the
discrepancy between numerical and analytical waveforms until the very
last few orbits, where ``tidal'' effects and missing high-order 
post-Newtonian coefficients also play a significant role.

We test different methods for extrapolating the gravitational wave
signal extracted at finite radii to null infinity. We show that
an effective procedure for integrating the Newman-Penrose $\psi_4$ signal
to obtain the gravitational wave strain $h$ is to apply a simple
high-pass digital filter to $h$ after a time domain integration, 
where only the two physical motivated integration constants are introduced.
That should be preferred to the more common procedures of introducing additional
integration constants, integrating in the frequency domain  or
 filtering $\psi_4$ before integration.
\end{abstract}


\vspace{2pc}

\submitto{\CQG}

\section{Introduction}
\label{sec:intro}

The recent, first, direct detection
\cite{Abbott:2016blz} of gravitational waves (GW) from a binary black
hole merger by Advanced Ligo \cite{TheLIGOScientific:2014jea} has
opened a new window for the investigation of astrophysical compact
objects. The new generation GW detectors Advanced LIGO and Advanced
Virgo \cite{TheVirgo:2014hva} are also expected to reveal an incoming
gravitational transient signal from Binary Neutron Stars (BNS)
coalescence and merger, once their sensitivity at higher frequencies
will increase. At design sensitivity, the rate of BNS signals detected
is predicted to be in the interval (0.2-200) per year \cite{LIGOVIRGO:2013},
making them the next target for GW detection. These expected detections present
a unique way to learn about the physics of matter at the extreme conditions
present in neutron stars and the EOS of nuclear matter above the nuclear
density.

Fully General Relativistic simulations of BNS started in 1999 \cite{Shibata:1999wm}
but it is since the crucial breakthroughs of 2005 \cite{Pretorius:2005gq,Campanelli:2005dd,Baker:2005vv} 
that numerical relativity is the main instrument to study the dynamics of the 
merger of compact objects. This has lead to the development of community driven 
public software like the Einstein Toolkit \cite{EinsteinToolkit:web,Loffler:2011ay,Moesta:2013dna}
and the LORENE library \cite{lorene:web,Gourgoulhon:2000nn} that allow to openly 
simulate such systems and in particular BNS coalescence and merger \cite{DePietri:2015lya}.
Simulation of BNS mergers,  where numerical relativity is almost the only available 
tool, is one of the main endeavors to study the possible
gravitational waveforms that will be discovered by the present detectors. For this 
reason there is now a considerable effort (using both public and private codes)  in analyzing 
the effect of the EOS used to describe matter on the gravitational-wave  signal (see
\cite{read:2013matter,Hotokezaka:2013mm,Hotokezaka:2015xka,Baiotti:2011am,bauswein:2015unified,Bauswein:2015vxa,Takami:2014tva,kastaun:2015properties,bauswein:2014revealing,hotokezaka:2013remnant,Lehner:2016lxy}
and reference therein), and to explore  other directions like incorporating in the codes more microphysical ingredients, such as finite temperature
nuclear EOSs with neutrino emission
\cite{kastaun:2015properties,Foucart:2015gaa,Bernuzzi:2015opx,Palenzuela:2015dqa,Sekiguchi:2015dma,Sekiguchi:2016bjd}
and magnetic fields
\cite{Palenzuela:2015dqa,Kiuchi:2015sga,Dionysopoulou:2015tda,Ponce:2014sza,Giacomazzo:2014qba,Kiuchi:2015qua,Palenzuela:2013hu,Rezzolla:2011da},
to accurately simulate the physics of the post-merger remnant (either
a (hyper)massive neutron star (HMNS) or a black hole (BH) surrounded
by an accretion disk).

However, despite all the progress, numerical modeling of BNSs should still be
considered an exploratory study due to a number of not completely settled
issues. Some of those issues are: the techniques for initial data generation and
the determination of their accuracy; the evolution code convergence properties
and the selection of the optimal numerical resolution; and, more importantly,
the choice of an appropriate modeling for the neutron star
EOS, still unknown from the microphysical and the experimental points of view.

This is reflected by the fact that there is also a lot of effort in trying 
to develop analytical techniques to compute the
gravitational-wave signal (to be compared and validated by numerical relativity ones) 
from the inspiral phase of compact binary mergers \cite{Boyle:2007ft,Blanchet:2013haa}, 
among which the effective one body (EOB)
\cite{Damour2008,buonanno:1999effective,Buonanno:2007pf} approach was
particularly successful. Recently, tidal effects
have been included in those analytical models
\cite{damour:2010effective,Hinderer:2016eia,Hotokezaka:2013mm,Bernuzzi:2014owa}. From
the measurements of tidal effects in the inspiral gravitational signal
the EOS of the neutron star core can be distinguished
\cite{Hotokezaka:2013mm,Hotokezaka:2015xka,Bernuzzi:2012ci,Baiotti:2011am,Hotokezaka:2016bzh,read:2013matter}. 
To contribute to this task, many-orbits (>10) BNS
simulations with nuclear EOSs have been performed lately, comparing
their gravitational signals to various EOB formulations with tidal
effects and Taylor-expanded post-Newtonian (PN) expressions
\cite{Hotokezaka:2015xka,Hotokezaka:2016bzh,Bernuzzi:2014owa,Haas:2016cop}.
The latter is challenging, because most of the initial
data used are eccentric, and the number of orbits used are not, in any sense, 
close to the $\simeq$ 175 orbits that were used in similar studies of 
binary black hole systems~\cite{Szilagyi:2015rwa}.

The present work improves the results
of~\cite{DePietri:2015lya} focusing on the techniques to extract the
gravitational-wave signal from numerical simulations; the analysis of the
eccentricity of the orbits obtained evolving quasi-circular initial
data and its impact on the gravitational waveforms;
the dependence of results on the number of simulated pre-merger orbits, and
different neutron star EOSs. 

In detail, we present new long-term (up to 16 orbits) equal-mass BNS simulations
with four different nuclear EOSs, starting with four different values
of the interbinary distance $d$ ($40$, $44.3$, $50$, and $60$ km). The
comparison of simulations starting from different initial orbital
frequencies is a necessary and fundamental test to evaluate the accuracy
of current numerical BNS simulations and their ability to model tidal
effects. In particular, long numerical simulations are needed to
construct hybrid analytical-numerical waveforms for GW detectors data
analysis where it is important to know how many orbits
before merger can be effectively simulated using current numerical 
methods and resolutions.

The outline of this paper is the following: in section \ref{sec:setup}
we describe the numerical methods employed in our code and the initial
models we choose to evolve. In section \ref{sec:gw} we describe the
techniques we tested for extracting, integrating and extrapolating to
null infinity the gravitational-wave signal. In section
\ref{sec:eccentricity} we briefly comment on the eccentricity of the
orbits from the evolution of quasi-circular initial data. In section
\ref{sec:comparison} we present our results about the comparison of
simulations of the same model starting with different initial
frequencies.  Finally, in section \ref{sec:conclusion} we summarize the
main results of the present work.

Throughout this paper we use a space-like signature $-,+,+,+$, with
Greek indices running from 0 to 3, Latin indices from 1 to 3, and the
standard convention for summation over repeated indices. The
computations are performed using the standard $3+1$ split into
(usually) space-like coordinates $(x,y,z)=x^i$ and a time-like
coordinate $t$. Our coordinate system $(x^\mu)=(t,x^i)=(t,x,y,z)$
(far from the origin) has, as it can be checked, almost isotropic
coordinates and far from the origin they would have the usual measure
unit of ``time'' and ``space''. In particular, $t$ is time when
measured from an observer at infinity. All computations have been done in normalized computational units
(hereafter denoted as CU) in which $c=G=M_\odot=1$. We report the
radius of the spheres used for gravitational waves extraction in
CUs.

\section{Numerical Methods and Initial Models}
\label{sec:setup}

We used the exact same code setup as our previous study about binary neutron star mergers \cite{DePietri:2015lya}, which is why we refer the interest reader to that work and the references within for details of the setup. Here we only summarize for clarity the most important components and numerical algorithms. A fundamental aspect for us is that this work has been produced using only free, open source software and as such can be reproduced and extended by anyone in the scientific community.

The dynamical evolution of the analyzed models was computed using the Einstein Toolkit \cite{EinsteinToolkit:web,Loffler:2011ay,Moesta:2013dna}. The Einstein Toolkit is a free, publicly available, community-driven general relativistic (GR) code based on the \codename{Cactus} \cite{Cactuscode:web,Goodale:2002a} computational framework. In particular, we have chosen the eleventh release (code name ``Hilbert'', ET\_2015\_05). Some local modifications and additions were necessary, all of which are open-source and freely available from the Subversion server of the Parma University gravity group \cite{SVN:2016}, and all of which are planned to be proposed for the next release of the Einstein Toolkit.

The data are evolved on a Cartesian mesh with 6 levels of mesh refinement, each with a double resolution respect to the parent level. The outer boundary of the grid is set at $1063$ km from its center, to be able to extract the gravitational signal far from the source. The standard grid spacing on the innermost level used for this paper is $dx = 369$ m, which in \cite{DePietri:2015lya} we found sufficient to get qualitatively good results. We used a mirror symmetry across the $(x,y)$ plane consistent with the symmetry of the problem, in order to reduce the computational costs by a factor two. The grid structure and the code parallelization are handled by the \codename{Carpet} code \cite{CarpetCode:web,Schnetter:2003rb}.

The spacetime metric evolution is performed by the \codename{McLachlan} module \cite{McLachlan:web}, implementing a $3+1$ dimensional split of the Einstein Equations using the BSSN-NOK formalism \cite{Nakamura:1987zz,Shibata:1995we,Baumgarte:1998te,Alcubierre:2000xu,Alcubierre:2002kk} and solving them with fourth-order finite differences.
The General Relativistic Hydrodynamics (GRHD) equations are solved by the module \codename{GRHydro} \cite{Moesta:2013dna} with High Resolution Shock Capturing techniques using the fifth-order WENO reconstruction method \cite{Harten:1987un,Shu:1999ho} and the HLLE Riemann solver \cite{Harten:1983on,Einfeldt:1988og}.
The time integration is performed with a fourth-order Runge-Kutta method \cite{Runge:1895aa,Kutta:1901aa} with a constant Courant factor of $0.25$. Kreiss-Olinger dissipation \cite{Kreiss:1973aa} is applied to the curvature evolution quantities in order to damp high-frequency noise.

We chose in particular the combination of the BSSN-NOK formulation of the Einstein equations and the WENO reconstruction method among the possibilities provided by the Einstein Toolkit because in \cite{DePietri:2015lya} we observed that combination of algorithms to enter the convergent regime already at low resolutions, allowing us to obtain reliable results without the need to perform computationally expensive simulations at higher resolution.

Our initial data are generated with the open source LORENE library \cite{lorene:web,Gourgoulhon:2000nn}, which is able to compute quasi-equilibrium configurations for binary irrotational neutron stars in quasi-circular orbits. This is achieved imposing the presence of the helical Killing vector $l^{\alpha} = \left(\partial_t\right)^{\alpha} +  \Omega \left( \partial_{\phi} \right)^{\alpha}$. From the imposition of this symmetry the initial data do not incorporate a radial velocity component coming from the gravitational radiation reaction. This will lead to the presence of a tangible eccentricity in the stars orbit during their evolution (see section \ref{sec:eccentricity} and ref. \cite{Dietrich:2015pxa,Kyutoku:2014yba} for more details).

\subsection{Equation of State}
The true EOS for nuclear matter in an environment similar to a neutron star is still not known,
not even assuming a small effect of the temperature, i.e., a cold neutron star, as expected here
for initial data. We thus have to simulate the effect of different plausible EOSs on observable
quantities, in the hope to learn about the EOS indirectly through these observations.

The four cold EOSs for nuclear matter at beta equilibrium that we employed are, in decreasing order of compactness:
\begin{itemize}
\item the \textbf{APR4} EOS \cite{Akmal:1998cf}, obtained using variational chain summation methods with the Argonne two-nucleon interaction and including also boost corrections and three-nucleon interactions,
\item the \textbf{SLy} EOS \cite{Douchin00,Douchin01}, based on the Skyrme Lyon effective nuclear interaction,
\item the \textbf{H4} EOS \cite{Lackey:2005tk}, constructed in a relativistic mean field framework including also Hyperons contributions and tuning the parameters to have the stiffest possible EOS compatible with astrophysical data, and
\item the \textbf{MS1} EOS \cite{Muller:1995ji}, constructed with relativistic mean field theory considering only standard nuclear matter.
\end{itemize}
\begin{figure}
\begin{centering}
\begin{subfigure}[t]{.48 \textwidth}
  \includegraphics[width=\textwidth]{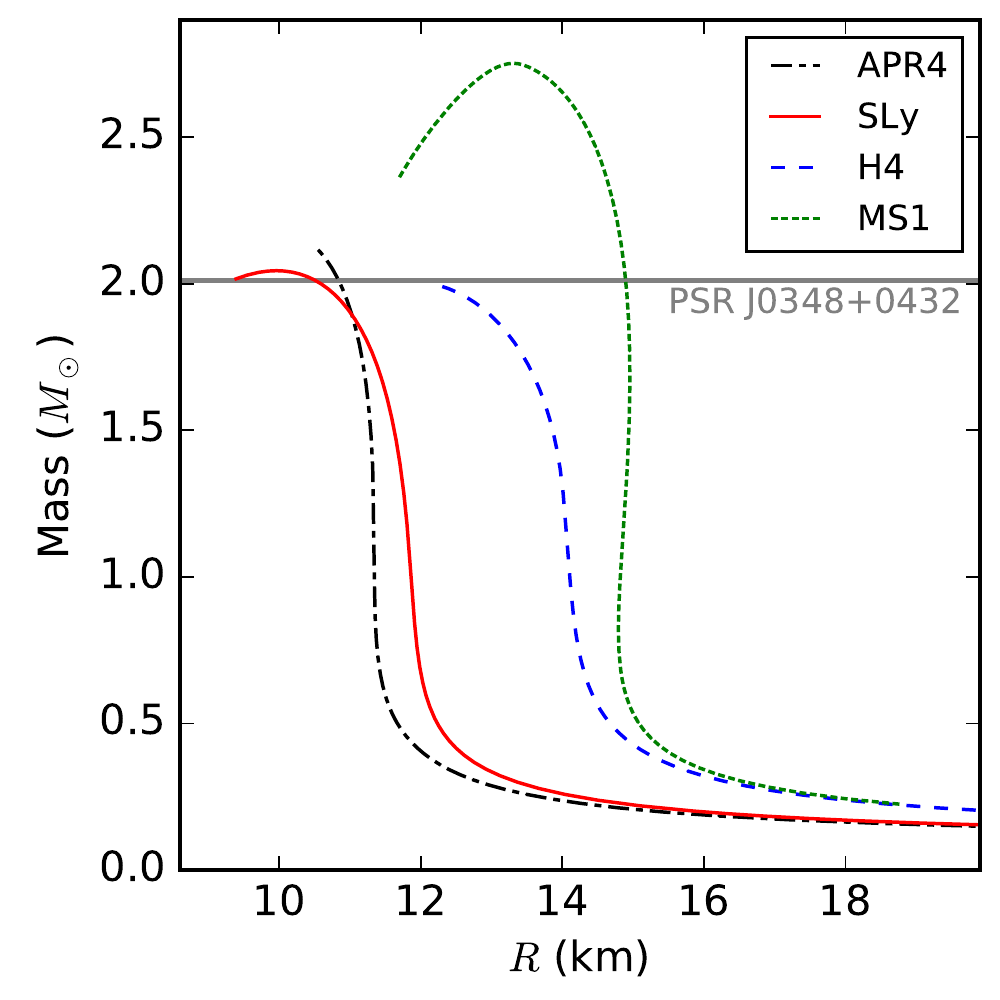}\\
\vspace{-6mm}
\caption{Mass-Radius relations}
\label{fig:eos}
\end{subfigure}%
\begin{subfigure}[t]{.48 \textwidth}
  \includegraphics[width=\textwidth]{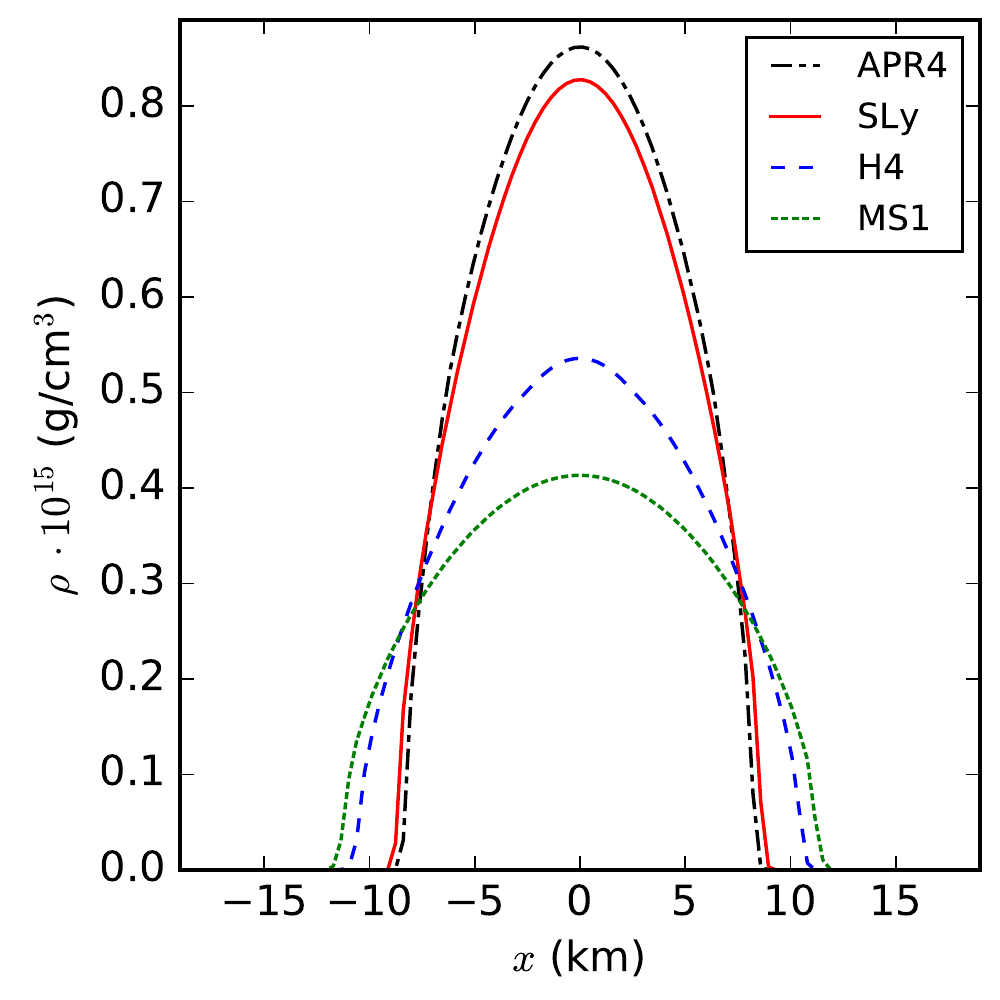}\\
\vspace{-6mm}
\caption{Initial density profiles}
\label{fig:in_dens}
\end{subfigure}%
\end{centering}
\caption{Left: Mass-radius relations for non-rotating neutron stars with four nuclear EOSs. The horizontal gray line marks the mass of PSRJ0348+0432 $M=2.01M_{\odot}$\cite{Demorest:2010bx}. 
Right: Initial density profiles for the stars using these four EOSs, with a Baryon mass of
$1.4M_\odot$.}
\end{figure}

Each of these EOSs satisfies the observational constraint of a maximum mass for a non rotating NS higher than $2.01 M_{\odot}$\cite{Demorest:2010bx}. In Fig.~\ref{fig:eos} and Fig.~\ref{fig:in_dens} we show the mass over radius curves and  the initial density profiles respectively, for non rotating neutron stars for each of these EOS. They have been generated solving the TOV equations with the code \codename{rns} \cite{Stergioulas95}.

We followed \cite{Read:2009constraints} parameterizing the EOSs as piecewise polytropes, with the following expressions in each density interval $\left[\rho_{i-1},\rho_i\right]$:
\begin{eqnarray} 
 P_{\mathrm{cold}}        &=& K_i \rho^{\Gamma_i} \\
 \epsilon_{\mathrm{cold}} &=& \epsilon_i + \frac{K_i}{\Gamma_i-1}\rho^{\Gamma_i-1} .
\end{eqnarray}
For each EOS we used 7 polytropic pieces, of which the first four (at lower densities) always adopt the prescription of \cite{Douchin00,Douchin01} for the stellar crust and the three remaining, instead, use the coefficient found in appendix B of \cite{Read:2009constraints} for each of the four EOS models described above. Characteristics of the employed EOSs and of their impact on the initial models are listed in Tab.~\ref{table:EOS}.\\
\begin{table}
\begin{tabular}{|c|ccc|c|cc|c|}
\hline
EOS & $\Gamma_4$ & $\Gamma_5$ & $\Gamma_6$ & $\rho_3 \left(g/cm^3\right)$ & $M_{grav} \left[M_{\odot}\right]$ & $R \left(km\right)$& C\\
\hline
APR4 & 2.830 & 3.445 & 3.348 & $1.512\times10^{14}$ & 1.2755 & 11.33 & 0.1662\\
\hline
SLy & 3.005 & 2.988 & 2.851 & $1.4622\times10^{14}$ & 1.2809 & 11.76 & 0.1608\\
\hline
H4 & 2.909 & 2.246 & 2.144 & $0.8877\times10^{14}$ & 1.3003 & 14.00 & 0.1371\\
\hline
MS1 & 3.224 & 3.033 & 1.325 & $0.9416\times10^{14}$ & 1.3048 & 14.90 & 0.1293\\
\hline
\end{tabular}
\caption{Characteristics of the employed EOSs and of the corresponding initial models. In the first three columns we report the adiabatic $\Gamma$ index of the three highest density polytropic pieces (starting with density $\rho_3$ and separated by densities $\rho_4=10^{14.7}g/cm^3$ and $\rho_5=10^{15}g/cm^3$) taken from \cite{Read:2009constraints}. In the fourth column we report the density that separates the crust EOS from the NS core EOS. The last three columns show: the gravitational mass at infinite separation, the radius and the compactness $C=M_{grav}/R$ of each star, respectively. }
\label{table:EOS}
\end{table}

During the evolution, the EOS is supplemented by a thermal component of the form
\begin{equation}
 P_{\mathrm{th}}          = \Gamma_{\mathrm{th}} \rho (\epsilon - \epsilon_{\mathrm{cold}}),
\end{equation}
choosing $\Gamma_{\mathrm{th}} = 1.8$ following the results of the discussion in \cite{bauswein:2010testing}.

\section{Results}
\label{sec:results}
We analyzed the dynamics of the inspiral phase of a BNS merger, performing numerical simulations of an equal mass binary system using four different EOSs for the cold nuclear matter and starting the dynamical evolution from four different values of coordinate distance between the star centers.

Since the gravitational mass of each star is not conserved during the evolution but depends on the gravitational binding energy we decided to generate our initial models fixing the conserved Baryonic mass to $1.4 M_{\odot}$. This leads to small differences in the gravitational masses at infinite separation for the different EOSs. Table \ref{table:InitialData} sums up the main physical characteristics of our initial models.

\begin{table}
\begin{tabular}{|c|cccc|}
\hline
Model name & $d_0 \left(km\right)$ & $M_{ADM}\left(M_{\odot}\right)$ & $J_{ADM}\left(G M_{\odot}^2/c\right)$ & $\Omega_0\left(rad/s\right)$\\
APR4(a) & 40 & 2.5255 & 6.3594 & 2036\\
APR4(b) & 44.3 & 2.5275 & 6.5771 & 1767\\
APR4(c) & 50 & 2.5298 & 6.8603 & 1492\\
APR4(d) & 60 & 2.5329 & 7.3385 & 1153\\
\hline
SLy(a) & 40 & 2.5361 & 6.4051 & 2040 \\
SLy(b) & 44.3 & 2.5382 & 6.6233 & 1770\\
SLy(c) & 50 & 2.5404 & 6.9073 & 1494\\
SLy(d) & 60 & 2.5435 & 7.3876 & 1155\\
\hline
H4(a) & 40 & 2.5744 & 6.5876 & 2056\\
H4(b) & 44.3 & 2.5764 & 6.8024 & 1782\\
H4(c) & 50 & 2.5787 & 7.0876 & 1504\\
H4(d) & 60 & 2.5819 & 7.5740 & 1163\\
\hline
MS1(a) & 40 & 2.5833 & 6.6434 & 2063 \\
MS1(b) & 44.3 & 2.5852 & 6.8503 & 1786\\
MS1(c) & 50 & 2.5875 & 7.1313 & 1506\\
MS1(d) & 60 & 2.5907 & 7.6164 & 1164\\
\hline
\end{tabular}
\caption{Initial binary systems parameters. Model names are listed using the following convention: the EOS name followed by a key indicating the initial  separation between the star centers (a: $40$ km, b: $44.3$ km, c: $50$ km, d: $60$ km). In this table we list the initial ADM mass of the systems, the initial total angular momentum and the initial orbital angular velocity, as reported by the output of LORENE.}
\label{table:InitialData}
\end{table}

Models with different EOS spend a different amount of time orbiting each other before they merge. This is mainly due to the different tidal deformability of the stars. However, some finite-resolution effect could be present, because simulations of models with different EOS can have different convergence properties, as found for example in~\cite{Hotokezaka:2015xka}. For this reason, merger time results from fixed, finite, resolution simulations must be taken with care. In Tab.~\ref{table:Orbits}, we report the number of orbits evolved in each simulation from the start to the merger.
\begin{table}
\begin{centering}
\begin{tabular}{|c||c|c|c|c|}
\hline
\multirow{2}{*}{EOS} & \multicolumn{4}{c|}{Number of orbits}\\
\cline{2-5}
& $d=60$ km & $d=50$ km & $d=44.3$ km & $d=40$ km\\
\hline
APR4 & 15.5 & 9.5 & 6.5 & 4.5\\
\hline
SLy  & 16 & 9 & 6 & 4 \\
\hline
H4   & 15 & 8 & 4.5 & 3\\
\hline
MS1  & 12.5 & 6 & 3.5 & 2 \\
\hline
\end{tabular}\\
\end{centering}
\caption{Number of orbits before merger, defined as the time for which the gravitational signal amplitude is maximum. We counted the orbits as half the number of gravitational wave cycles before the merger.
}
\label{table:Orbits}
\end{table}

In the following subsections we will first present details about the gravitational wave extraction procedure we implemented. Next, we analyze the eccentricity of the orbits obtained evolving the quasi-circular LORENE initial data, and evaluate its impact on the gravitational signal. Finally, we evaluate the effect of the choice of initial interbinary distance on the quality of the simulation results.

\subsection{Gravitational wave extraction}
\label{sec:gw}
We extracted the gravitational-wave signal from each simulation with the standard method of calculating during the numerical evolution the Newman-Penrose scalar $\psi_4$ \cite{Newman:1961qr,Baker:2001sf} (using to the Einstein Toolkit module \codename{WeylScal4}), decomposed in spin-weighted spherical harmonics of spin $-2$ \cite{Thorne:1980ru} by the module \codename{Multipole}:
\begin{equation}
 \psi_4(t,r,\theta,\phi)  = 
 \sum_{l=2}^{\infty} {\sum_{m=-l}^{l} {\psi_4^{lm}(t,r)\ {{}_{-2}\!}{Y}_{lm}(\theta,\phi)}}.
\end{equation}
We extracted $\psi_4$ components up to $l=6$. At null infinity, $\psi_4$ is linked to the gravitational waves strain by the relation \beq{\psi_4\ =\ \ddot{h}_+\ -\ i \ddot{h}_x\ :=\  \ddot{\bar{h}}, \label{eq:h}} where $\bar{h}$ stands for the complex conjugation of the GW strain. In order to integrate it twice to get the strain components $h^{lm}$, we used a simple trapezoidal rule, starting the integration from coordinate time $t=0$, as suggested by~\cite{Damour2008}. We performed a polynomial fit to the obtained strain and subtracted its result to the strain itself. In order to correctly set the integration constants a linear fit is used:
\begin{eqnarray}
\bar{h}^{(0)}_{lm}\ &=&\ \int_{0}^t{dt' \int_{0}^{t'}{dt'' \psi_4^{lm}(t'',r)}} \label{eq:time_int}\\
\bar{h}_{lm}\ &=&\ \bar{h}^{(0)}_{lm}\ -\ Q_1 t\ -\ Q_0.\label{eq:fit_int}
\end{eqnarray}
Like in our previous work \cite{DePietri:2015lya}, however, we found also here that the linear fit is not sufficient to eliminate the low-frequency unphysical oscillations in the strain amplitude, caused by unresolved high-frequency noise aliased in the low-frequency signal during the integration process~\cite{Reisswig:2011notes}.
In~\cite{DePietri:2015lya}, we found that subtracting from the strain a second order polynomial fit is sufficient to eliminate the unphysical drift from the dominant $(2,2)$ mode, which is the one we focus on in the rest of this paper unless otherwise stated.

Unfortunately, this procedure is not able to cure the nonphysical oscillations in the sub-dominant modes, for which fitting with even higher order polynomials would be needed, as first recognized in \cite{Berti:2007inspiral}. Since some analysis presented later in this paper are based on the emitted angular momentum, for which the contribution of sub-dominant modes is relevant (see Eq. (\ref{eq:J})), we found that the standard procedure also used in \cite{DePietri:2015lya} need to be improved since the use of higher-order fitting polynomials is not justified from the mathematical point of view. Here we discuss differents approaches to the problem and propose a new one.

An almost universally applicable integration procedure is the so-called Fixed Frequency Integration (FFI), introduced in~\cite{Reisswig:2011notes}. It consists of integrating the signal in the frequency domain with the following prescription:
\begin{eqnarray}
\tilde{h}_{lm}(f) = -\frac{\tilde{\Psi}_4^{lm}(f)}{\left(2\pi f_0\right)^2}\ \ &\mbox{if}&\ \ f < \frac{m f_0}{2} \label{eq:ffi-}\\
\tilde{h}_{lm}(f) = -\frac{\tilde{\Psi}_4^{lm}(f)}{\left(2\pi f\right)^2}\ \  &\mbox{if}&\ \ f \geq \frac{m f_0}{2}, \label{eq:ffi+}
\end{eqnarray}
where $\tilde{}$ stands for the Fourier transform of a function and $f_0$ is the lowest physical frequency of the gravitational radiation emitted during the system evolution. This method is equivalent to applying a high-pass filter to $\Psi_{lm}$, damping the part of the signal spectrum below the frequency $\frac{m f_0}{2}$, by multiplying $\Psi_{lm}$ with a transfer function $H(f) = \frac{f^2}{f_0^2}$.
While satisfactory for most applications, we found two shortcomings in this integration procedure:
\begin{enumerate}
\item After applying the extrapolation formula of~\cite{Nakano:2015perturbative} (see Sec.~\ref{sec:extrapolation}) to $\psi_4$, the filter introduced by FFI is not strong enough to reduce the low frequency part of the signal sufficiently to avoid any visible artifact in the waves amplitude. This is due to the amplification of the low frequency components of $\psi_4$ by the additional integral terms in eq. \ref{eq:secondorder}. A possible strategy to overcome this problem is to increase the filter order, for example by changing the transfer function to $H(f) = \frac{f^4}{f_0^4}$.
\item Performing a direct and an inverse Fourier transform to obtain the time-domain double-integrated signal needs the application of a sufficiently smooth  window function to $\psi_4(t)$ before the integration to avoid problems due to the finite signal length. This means that a relevant fraction of the computed signal is not available for further analysis, requiring longer simulations, wasting computational resources.
\end{enumerate}
\begin{figure}
\begin{centering}
\includegraphics[width=0.95\textwidth]{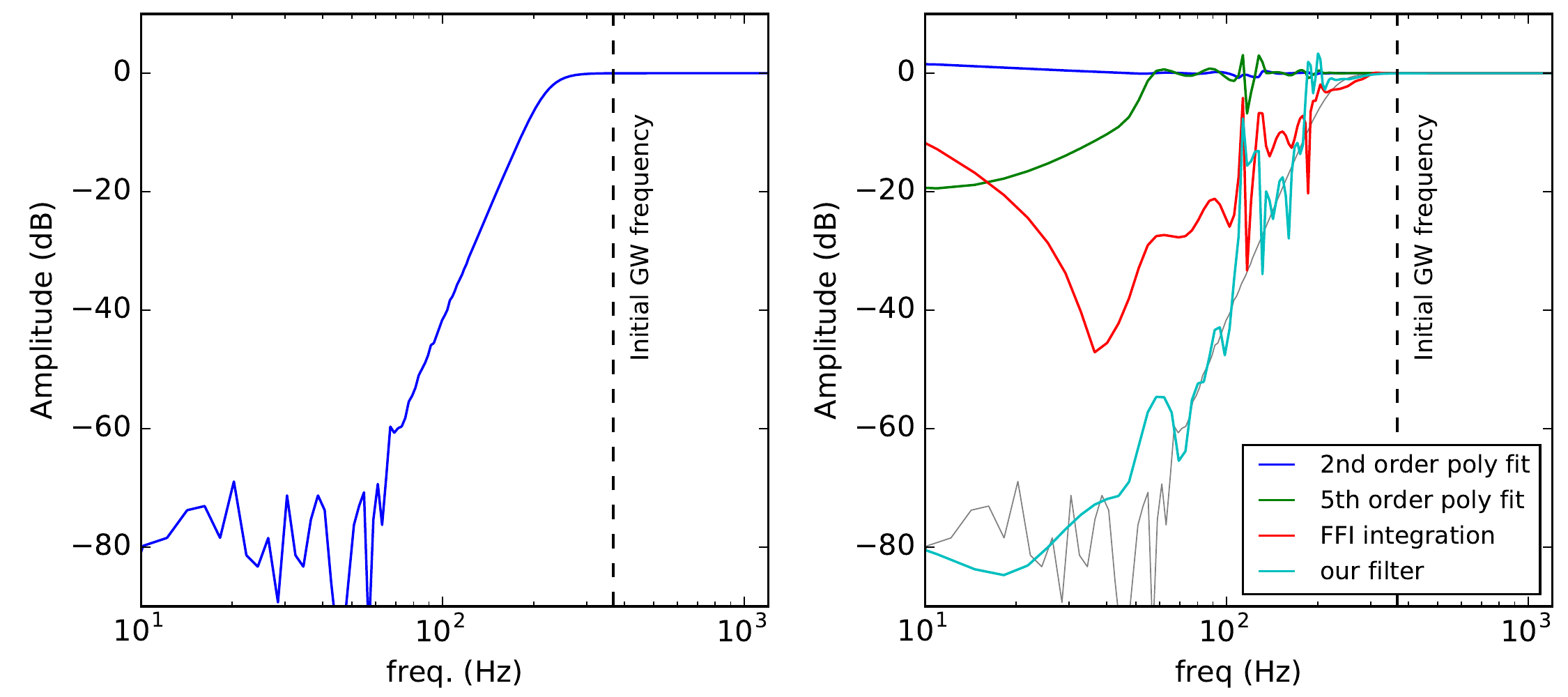}\\
\end{centering}
\caption{Left: the frequency response of the high-pass Butterworth filter applied to the gravitational wave strain to damp the unphysical amplitude oscillations. Right: the low frequency amplitude suppression caused by the different integration procedures used to compute $h$ form $\psi_4$, with respect to the strain calculated with Eq.~(\ref{eq:fit_int}) (see text for details). This plot refers to the model SLy(d). Before integration, $\psi_4$ has been extrapolated to null infinity with formula \ref{eq:secondorder} (see Sec.~\ref{sec:extrapolation}). }
\label{fig:integration}
\end{figure}

To overcome these limitations, we integrated $\psi_4$ in the time domain, subtracting a linear fit to fix the two integration constants, following equations \ref{eq:time_int} and \ref{eq:fit_int} (the correct procedure in absence of noise aliasing problems). Only after the integration we applied a digital high-pass Butterworth filter to $h(t)$, designed with the \codename{signal} module of the \codename{scipy} Python library. We created an IIR filter imposing a maximum signal suppression of $-0.01dB$ at the minimum physical frequency $f_0$ (computed as twice the initial orbital frequency) and a signal suppression of $-80dB$ at $\frac{f_0}{10}$, and applied it as a forward-backward filter to avoid changes to the signal phase. A filter applied to $h$ and not to $\psi_4$ requires a lower order, since the low-frequency noise components of $\psi_4$ are greatly amplified by the double integration process. In Fig.~\ref{fig:integration}, we show the frequency response of the applied filter and report on the frequency behavior $H(f)$ of the effective transfer functions corresponding to different integration procedures (subtracting different order polynomial fits, FFI integration and the use of our filter), expressed in decibels. They are calculated using the GW data of our simulations, with the following expression:
\begin{equation}
H_i(f) = 20\cdot \log_{10}\left|\frac{\tilde{h}_i{f}}{\tilde{h}_0(f)}\right|, \label{eq:dB}
\end{equation}
where $\tilde{h}_i(f)$ are the Fourier transforms of the GW strains computed with the different integration procedures and $\tilde{h}_0(f)$ is the Fourier transform of the GW strain computed with eq. \ref{eq:fit_int}. All the strains are computed starting from the second order extrapolated $\psi_4$ of eq. \ref{eq:secondorder}. It is clearly visible that subtracting a second order polynomial fit as we did in \cite{DePietri:2015lya} is no longer sufficient and that our filter performs better than FFI integration, in particular at the lowest frequencies.\\
After having obtained a clean $h(t)$ signal with our filter procedure described above, we were able to calculate the radiated energy and angular momentum fluxes \cite{Brugmann:2008calibration}:
\begin{eqnarray}
\frac{dE^{gw}}{dt}   &=& \frac{R^2}{16\pi}\int{d\Omega \left|\dot{h}(t,\theta,\phi)\right|^2} \\\label{EQ:dEdt}
\frac{dJ_z^{gw}}{dt} &=& \frac{R^2}{16\pi} Re \!\left[\int{\!\!d\Omega \label{EQ:dJdt}
       \left( \partial_{\phi}\; \dot{\bar{h}}(t,\theta,\phi) \right)
       h(t,\theta,\phi)}\right], \label{eq:J}
\end{eqnarray}
where in this case \beq{h(t,\theta,\phi) = \sum_{l=0}^6{\sum_{m=-l}^{l}{h_{lm}(t) {{}_{-2}\!}{Y}_{lm}(\theta,\phi)}}}
\subsubsection{Gravitational signal extrapolation to null infinity}
\label{sec:extrapolation}
We already noted that the relationship between $h$ and $\psi_4$ (Eq.~\ref{eq:h}) is valid only at null infinity. The extraction of the gravitational signal at finite radius will therefore introduce a source of error in the computed GW strain.

Different procedures have been developed to extrapolate the $\psi_4$ signal computed during the numerical evolution to null infinity. The three most important ones are:
\begin{itemize}
\item Extracting the signal at different radii and computing a fit to a polynomial in $\frac{1}{R}$ \cite{Boyle:2009vi};
\item Propagating the signal obtained at finite radius to null infinity with an analytic formula based on the results of perturbation theory of the Schwarzschild (or Kerr) spacetime \cite{Lousto:2010qx,Nakano:2015perturbative};
\item Performing a characteristic evolution using as inner boundary conditions the metric and its derivatives on a timelike worldtube (CCE) \cite{Bishop:1998uk,Babiuc:2010ze}.
\end{itemize}

In this work we confront the first two possibilities, extracting $\psi_4$ at seven different equidistant radii from $R=400$ CU ($591$ km) to $R=700$ CU ($1034$ km).\\
For the polynomial extrapolation we calculated the phase $\phi(t_\mathrm{ret})$ and amplitude $A(t_\mathrm{ret})$ of $\psi_4(t_\mathrm{ret}) = A(t_\mathrm{ret})e^{i\phi(t_\mathrm{ret})}$, where from now on we will express all wave functions with respect to the retarded time $t_\mathrm{ret}$: 
\begin{eqnarray}
t_\mathrm{ret} &=& t - R^*\\
R^* &=& R + 2M_\mathrm{ADM} \log\left(\frac{R}{2M_\mathrm{ADM}} - 1\right),
\end{eqnarray}
where $t$ is the coordinate time and $R$ is the radial coordinate of the extraction sphere in our numerical relativity coordinate system, which far from the stars is similar to isotropic coordinates.

Next, we fit both the amplitude and the phase of the signal extracted at different radii with a second order $\frac{1}{R}$ polynomial:
\begin{equation}
f(R,t_\mathrm{ret})\ = \ a_0(t_\mathrm{ret}) + \frac{a_1(t_\mathrm{ret})}{R} + \frac{a_2(t_\mathrm{ret})}{R^2},
\label{eq:polyex}
\end{equation}
where $f(R,t_\mathrm{ret})$ stands for $R\cdot A(R,t_\mathrm{ret})$ or $\phi(R,t_\mathrm{ret})$ and this fit is repeated for each discrete point in the time evolution of the system. The coefficient $a_0(t_\mathrm{ret})$ gives, for each time, the value of the function $f$ extrapolated to infinity. We extracted the gravitational signal at too few radii to have significant results with a higher order fit.\\
For the perturbative extrapolation, instead, we tested for the first time in a BNS simulation the second order correction introduced in \cite{Nakano:2015perturbative}, to see if it makes any substantial difference from the first order one already proposed in \cite{Lousto:2010qx} and tested in \cite{Lousto:2013oza,Hinder:2013oqa,Hotokezaka:2015xka}. This is also useful to have a possible estimation of the error due to finite radius extrapolation, that according to \cite{Chu:2015kft} is the dominant source of error in numerical waveforms in the early inspiral, far from the plunge phase.\\
The second order perturbative extrapolation formula is:
\begin{equation*}
r\psi_4^{lm}(t_\mathrm{ret})\left|_{r=\infty}\right. \ =\ \left( 1-\frac{2M}{r}\right) \left( r\psi_4^{lm}(t_\mathrm{ret},r) \right.
\end{equation*}
\begin{equation}
- \ \frac{(l-1)(l+2)}{2r}\int{dt\ r\psi_4^{lm}(t_\mathrm{ret},r)}
\label{eq:secondorder}
\end{equation}
\begin{equation*}
\left.\ +\ \frac{(l-1)(l+2)(l^2+l-4)}{8r^2}\int{dt^{'}\int{dt\ r\psi_4^{lm}(t_\mathrm{ret},r)}}\right).
\end{equation*}
where $r$ is the Schwarzschild radial coordinate:
\begin{equation}
r\ =\ R\left(1+\frac{M_{ADM}}{2R}\right)^2.
\end{equation}
The first coefficient $\left( 1-\frac{2M}{r}\right)$ comes from the difference between the `psikadelia' tetrad \cite{Baker:2001sf} commonly used in numerical relativity to compute $\psi_4$ and the Kinnersley tetrad \cite{Kinnersley:1969zza} used in perturbation theory.
Since the authors of \cite{Nakano:2015perturbative} do not see a strong dependence on (the Kerr parameter) a, we decided not apply the background spin corrections \cite{Nakano:2015perturbative}.\\
In practice, the integral terms in Eq.~(\ref{eq:secondorder}) needs to be treated with the same procedure used for obtaining $h$ from $\psi_4$ in order to eliminate any unphysical drift. To be consistent with the integration constants and not to mix filtered and unfiltered terms, we applied Eq. \ref{eq:secondorder} in the following form:
\begin{equation*}
r\psi_4^{lm}(t_\mathrm{ret})\left|_{r=\infty}\right. \ =\ \left( 1-\frac{2M}{r}\right) \left( r\ddot{\bar{h}}(t_\mathrm{ret},r) + \right.
\end{equation*}
\begin{equation}
\left.- \ \frac{(l-1)(l+2)}{2r} \dot{\bar{h}}(t_\mathrm{ret},r)\ +\ \frac{(l-1)(l+2)(l^2+l-4)}{8r^2} \bar{h}(t_\mathrm{ret},r)\right),
\end{equation}
where the gravitational wave strain $h$ is computed from $\psi_4$ extracted at finite radius $R$ with our filter procedure illustrated in section \ref{sec:gw}. The time derivatives are computed with a fourth order operator.\\
\begin{figure}[h]
\begin{centering}
  \includegraphics[width=\textwidth]{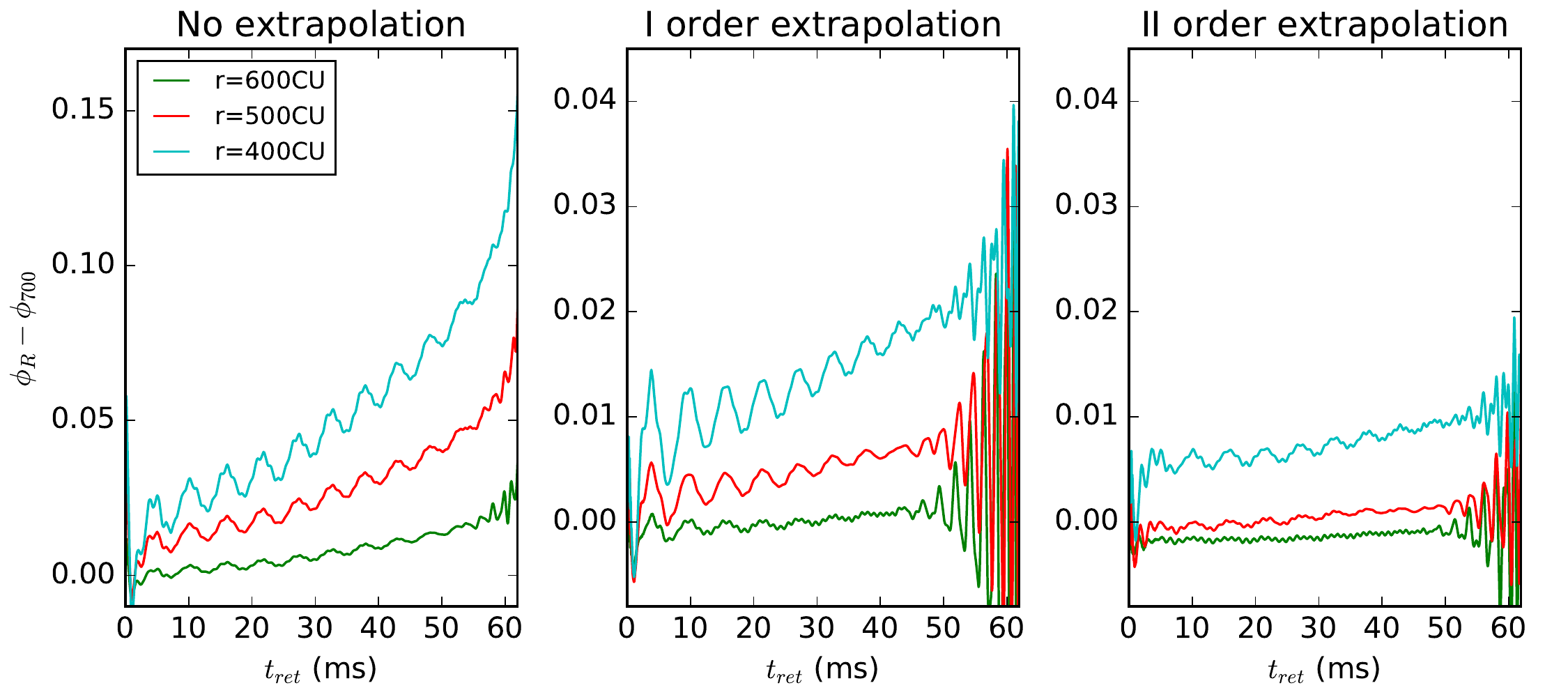}\\
\end{centering}
\vspace{-2mm}
\caption{Phase differences between the gravitational waves extracted at $R=400$ CU (cyan), $R=500$ CU (red), $R=600$ CU (green), and at $R=700$ CU. They are calculated for the example model SLy(d). The left panel are represents the original signals, the middle panel shows the perturbative extrapolated with the first order formula, and the right panel the perturbative extrapolated with the second order formula (eq. \ref{eq:secondorder}). The waves are aligned in phase at $t_\mathrm{ret}=0$. No further alignment procedure (like described in section \ref{sec:comparison}) is necessary because we are confronting waveforms extracted from the same simulation.}
\label{fig:phase_r}
\end{figure}
\begin{figure}
\begin{centering}
  \includegraphics[width=\textwidth]{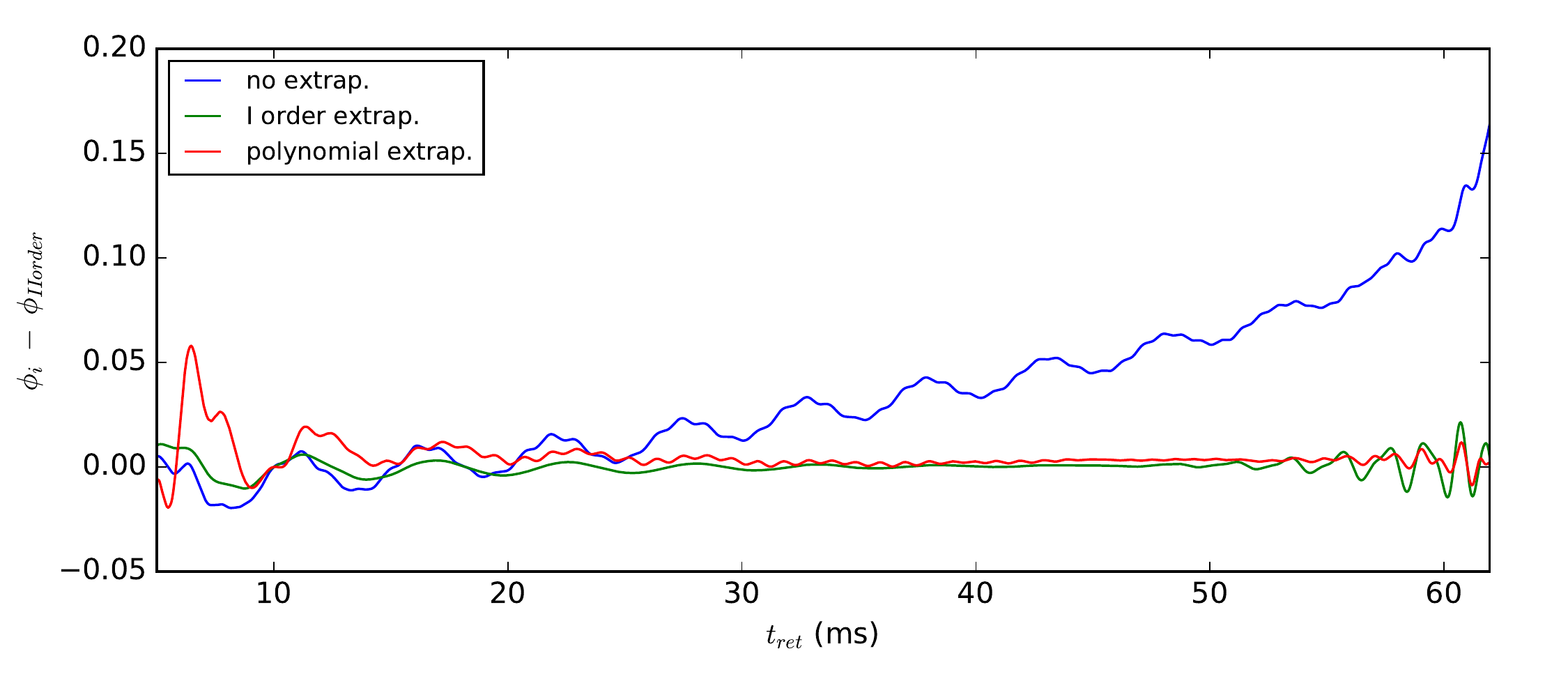}\\
\end{centering}
\vspace{-2mm}
\caption{Waveform phase differences using different extrapolation methods compared to the II order extrapolation.
Shown is data for the example model SLy(d). The waveforms are aligned at $t_\mathrm{ret}=10$ ms, to exclude phase differences from the initial spurious radiation, which would be dominant given the low magnitude of the phase differences between extrapolation methods.}
\label{fig:phase_methods}
\end{figure}

To evaluate the extrapolation formula effectiveness we computed, for the reference model SLy(d), the difference between the signal extracted at different radii, with and without applying the perturbative formula, and also compare the first order with the second order correction. The results of this comparison can be seen in Fig.~\ref{fig:phase_r}, where we show the difference of the wave strain phase $\phi(t) = \mathrm{atan}\left(\frac{h_{\times}(t)}{h_+(t)}\right)$, extracted at different radii with the one extracted at the outermost radius $R=700$ CU, for the three different extrapolation orders.

As can be seen in Fig.~\ref{fig:phase_r}, applying the first order perturbative correction to $\psi_4$ reduces the phase difference between the signal extracted at different radii by one order of magnitude. Applying also the second order correction reduces it further by about a factor two. These phase errors are consistent with the one shown in Fig.~\ref{fig:phase_methods}, where we can see that through most of the binary coalescence the difference between the perturbative extrapolated signal phase and the polynomial extrapolated signal phase is around $0.01$ rad and oscillates with a maximum under $0.05$ rad in the merger phase. Using this figure, we like to note that these errors are much smaller than the ones typically expected from finite-resolution effects for this kind of GRHD simulations, which, in the best cases reported in the literature, are of the order $0.5$ rad \cite{Hotokezaka:2015xka,Bernuzzi:2014owa,Hinder:2013oqa}, and of which we give a lower bound estimate in section \ref{sec:comparison}. Therefore, applying only the first order term in the perturbative extrapolation is sufficient if one is only interested in the waveform phase.

On the other hand, on the contrary  to what happen to the phase, we note that the difference in the emitted energy  
in gravitational waves  do not show any sensible difference on the extrapolation methods used or extraction radius. 
This is due to the fact that the correction in the gravitational wave amplitude starts at second order in $\frac{1}{R}$,
as shown  in \cite{Nakano:2015rda}. Indeed even if the difference between the signals extracted at different radii, is 
lower for the second order extrapolated signal with the respect to the one extract using  the first order extrapolated
signals they are  very small, with a variation of the order of tenths of percent and this is even for signal 
extracted at $r=700$ CU and $r=400$ CU without applying any extrapolation formula. We indeed 
report, for the rest of the present work, result obtained using the second order perturbative extrapolated signal (\ref{eq:secondorder}) using  
Eqs. (\ref{eq:time_int}) and (\ref{eq:fit_int}) and applying a
digital high-pass Butterworth filter. Moreover we denote as $t_{merger}$ the time where the $h_{22}$ amplitude 
as the maximum.

\subsection{Orbital Eccentricity}
\label{sec:eccentricity}
\begin{figure}
\begin{centering}
  \includegraphics[width=0.95\textwidth]{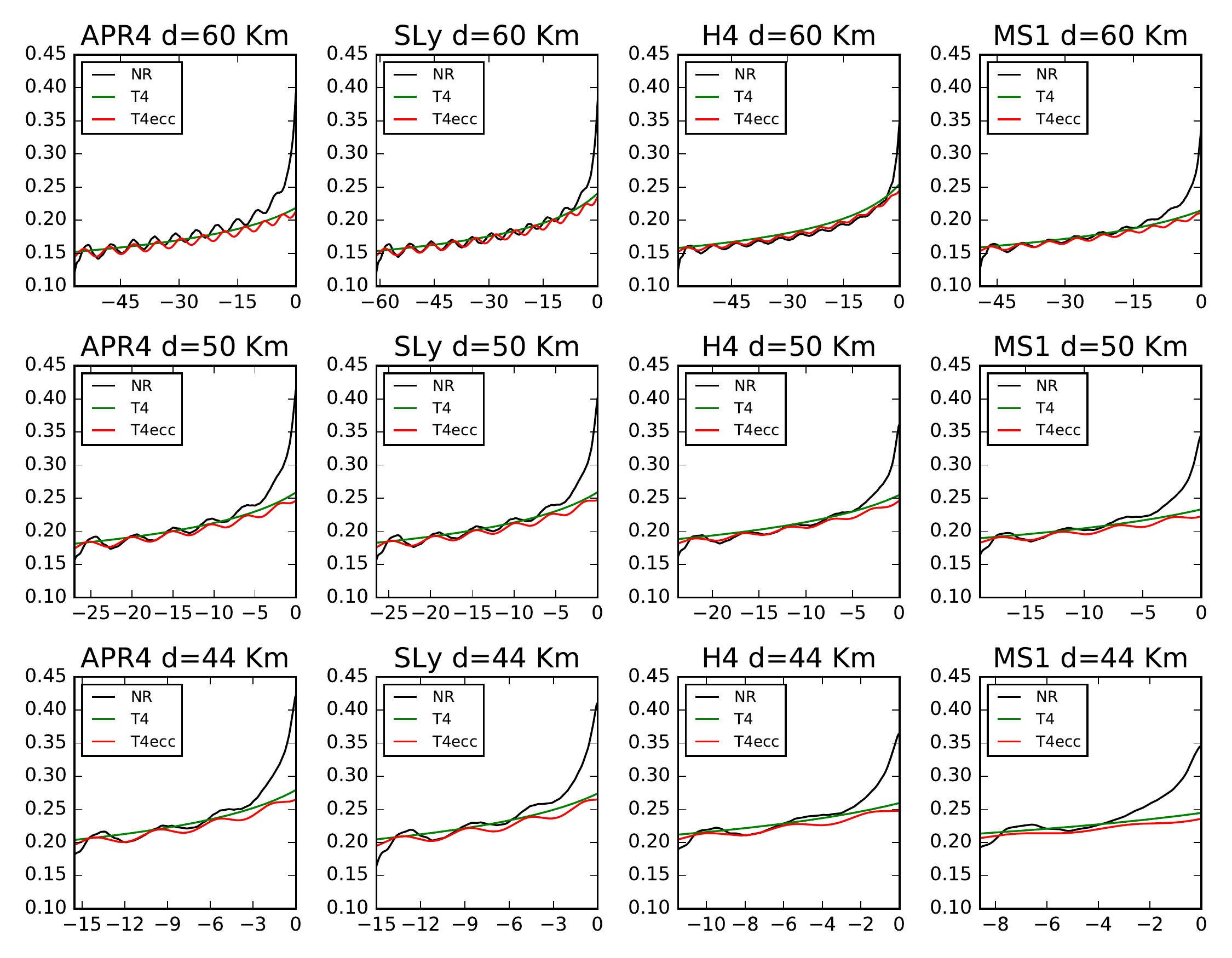}\\
\end{centering}
\vspace{-2mm}
\caption{Comparison of gravitational waves amplitudes of the simulated models with respected to two TaylorT4 post Newtonian approximations, with the same initial mass and frequency. The eT4 approximant includes also an initial orbital eccentricity equal to the one measured from the corresponding numerical simulation. These analytical waveforms do not include tidal effects, therefore agree with the numerical ones only in the first part of the signal and it should be noted that the main effect is connected to the eccentricity of the initial date except for the last few ms. Each box correspond to a different EOS and a different initial separation ($d$) of the Stars }
\label{fig:ecc}
\end{figure}

The amplitude of the gravitational wave strain shows a characteristic oscillation in all our simulations (see Figs.~\ref{fig:ecc} and \ref{fig:AllSims+}). This phenomenon has been interpreted in the literature as an imprint of a small eccentricity in the orbital evolution due to the missing approaching radial velocity in the quasi-circular initial data~\cite{Dietrich:2015pxa,Kyutoku:2014yba} (see Sec.~\ref{sec:setup}). We checked the trajectories of our BNS systems, calculating their eccentricity in a simple way, and later confronted their gravitational wave strain with the one computed with a recently developed analytical model for eccentric binaries~\cite{Tanay:2016zog}.

\begin{figure}
\begin{centering}
  \includegraphics[width=0.95\textwidth]{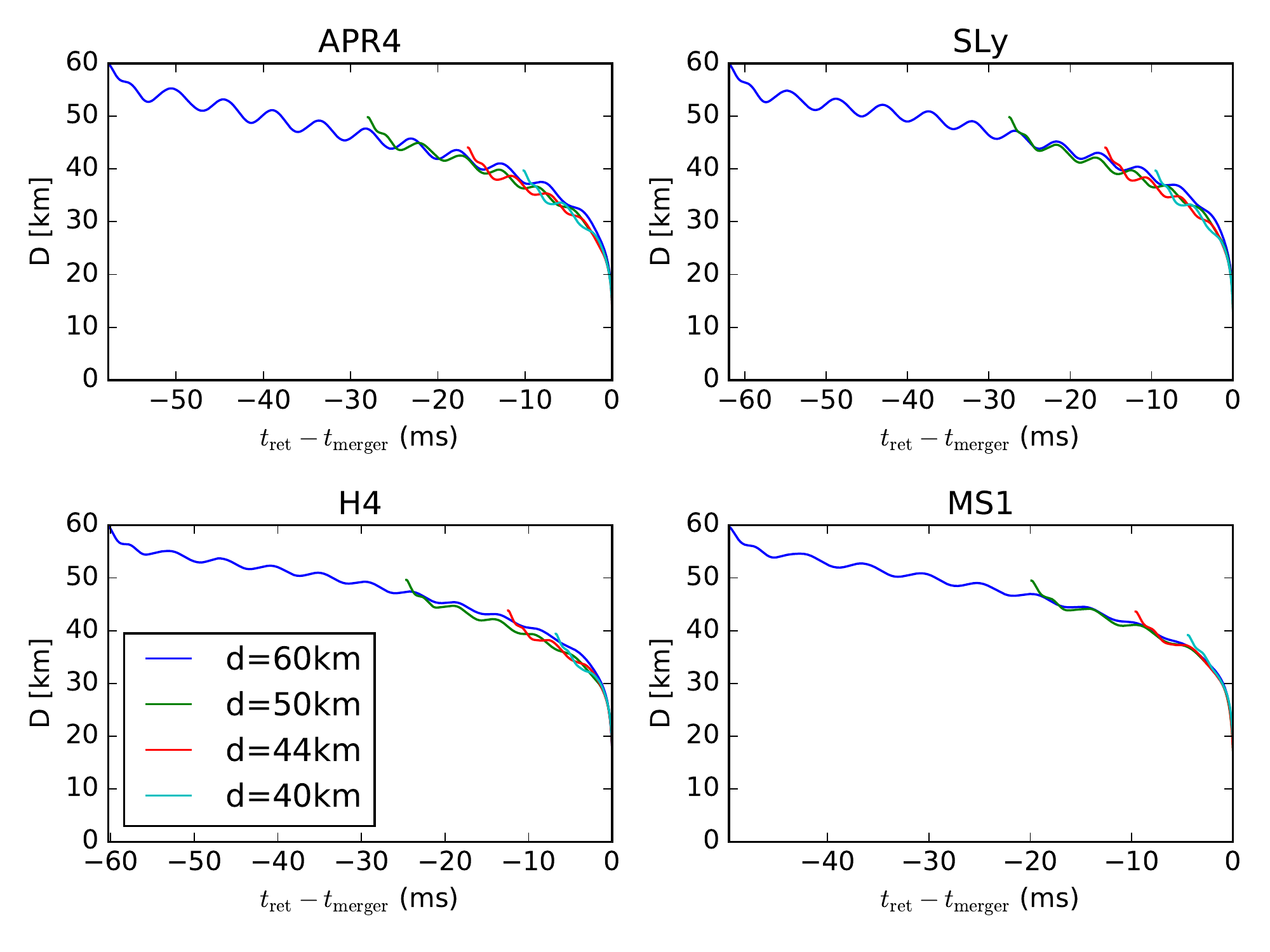}\\
\end{centering}
\vspace{-2mm}
\caption{Evolution of the coordinate distance between the star centers, assumed as the points with the maximum density inside each star. The effect of the orbital eccentricity is clearly recognizable in the distance oscillations. We note also that the orbital evolution of models starting at different initial distances does not perfectly match, in particular for the more compact models. Also see Sec.~\ref{sec:comparison} for a discussion.}
\label{fig:orbit}
\end{figure}
We computed the trajectories by following the dynamics of the star centers, defined as the points on the numerical grid with the maximum density $\rho$. Next, we computed the coordinate distance $D$ between the star centers at each time step, and fitted its derivative (computed with a fourth order operator) with the following Newtonian approximation for the orbital evolution:
\beq{\dot{D}(t)\ =\ A_0\ +\ A_1 t\ -\ e\ D_0\ \omega_e\ sin\left(\omega_e t + \phi_e\right)\label{eq:ecc},} 
where $e$ is the eccentricity and $D_0=d$ the initial coordinate interbinary distance. The fit is performed in the time interval between $t_\mathrm{ret}=3$ms and $t_\mathrm{ret}=\frac{2}{3} t_\mathrm{merger}$, to avoid the initial spurious radiation and the plunge phase but having at least one eccentricity cycle included. For the models starting from only $d=40$ km it is not possible to satisfy that last requirement, which is why we excluded those simulations from the analysis in this section.
We fitted the derivative of $D$ because of the advantage of having one free parameter less in the fit. We note that this way of defining an inter-star distance is not gauge independent, and also in no way unique. However, although the distance between the stars can be computed in more accurate ways, e.g., by finding their centers using an integral weighed with the density, in analog to a Newtonian center of mass, and by calculating the proper distance along a geodesic between the stars, the spacetime and matter quantities change only slowly during a long section of the inspiral, making as approximation as simple as ours viable.
\begin{table}
\begin{tabular}{|c|ccc||c|c|}
\hline
& e & e &e &  $\|h_{T4} - h_{eT4}\|$ & R [Mpc]\\
EOS  & $d=60$km& $d=50$km&$d=44.3$km & $R=100$Mpc & $\|h_{T4} - h_{eT4}\|=1$ \\
\hline
APR4 & 0.028 & 0.020 & 0.020 & 0.67 & 67\\
\hline
SLy & 0.025 & 0.019 & 0.020 & 0.58 & 58\\
\hline
H4 & 0.012 & 0.012 & 0.014 & 0.33 & 33\\
\hline
MS1 & 0.014 & 0.014 & 0.007 & 0.35 & 35\\
\hline
\end{tabular}
\caption{Results of the study about the orbital eccentricity in our simulations. The first three columns show the eccentricity parameters obtained fitting equation \ref{eq:ecc}. The fourth column shows the detectability of the eccentricity in a TaylorT4 approximate waveform with the same initial parameters of the model with the corresponding EOS and $d=60$~km, for an optimally aligned binary at $100$~Mpc. The fifth column shows the maximum optimally aligned binary distance for the eccentricity effect to be marginally detectable, calculated as $\|h_{T4} - h_{eT4}\|_{100Mpc}\times 100$~Mpc.}
\label{table:ecc}
\end{table}

In Tab.~\ref{table:ecc} we list the resulting values of initial orbital eccentricity for all the simulations with $d>40$ km. We cannot see any clear trend in the variation of the orbital eccentricity with different initial interbinary distances. We can see, instead, that the eccentricity for the more compact stars (with APR4 and SLy EOSs) is about two times the one for less compact stars (with H4 and MSI EOSs).

In the last years several groups developed a technique to produce eccentricity-reduced initial data for BNS simulations, through an iterative procedure \cite{Dietrich:2015pxa,Kyutoku:2014yba}. They start a simulation from quasi-circular data, calculating its orbital eccentricity. Then they compute the radial velocity which needs to be added to the initial data to correct the eccentricity effect and they repeat the procedure with the modified initial data, until the residual eccentricity is at least less than $2\times 10^{-3}$, as prescribed by the NRAR collaboration \cite{Hinder:2013oqa}.
We do not have yet implemented this algorithm in our codes. Instead, we want to evaluate the effect of the error induced by an orbital eccentricity of order $10^{-2}$, like the one we measure in this work.

In order to confirm that the measured eccentricity is indeed due to the non-ideal initial data,
we compare in Fig.~\ref{fig:ecc} our measured GW amplitude with both the standard T4~\cite{Boyle:2007ft} and the recently developed eccentric TaylorT4 approximant (eT4)~\cite{Tanay:2016zog}. For now their model is only accurate to the 2PN order in phase, to the Newtonian level in amplitude, and does not include tidal effects, but it is nevertheless in very good agreement with our numerical results, at least in the first part of the signal.

\begin{figure}
\begin{centering}
  \includegraphics[width=0.95\textwidth]{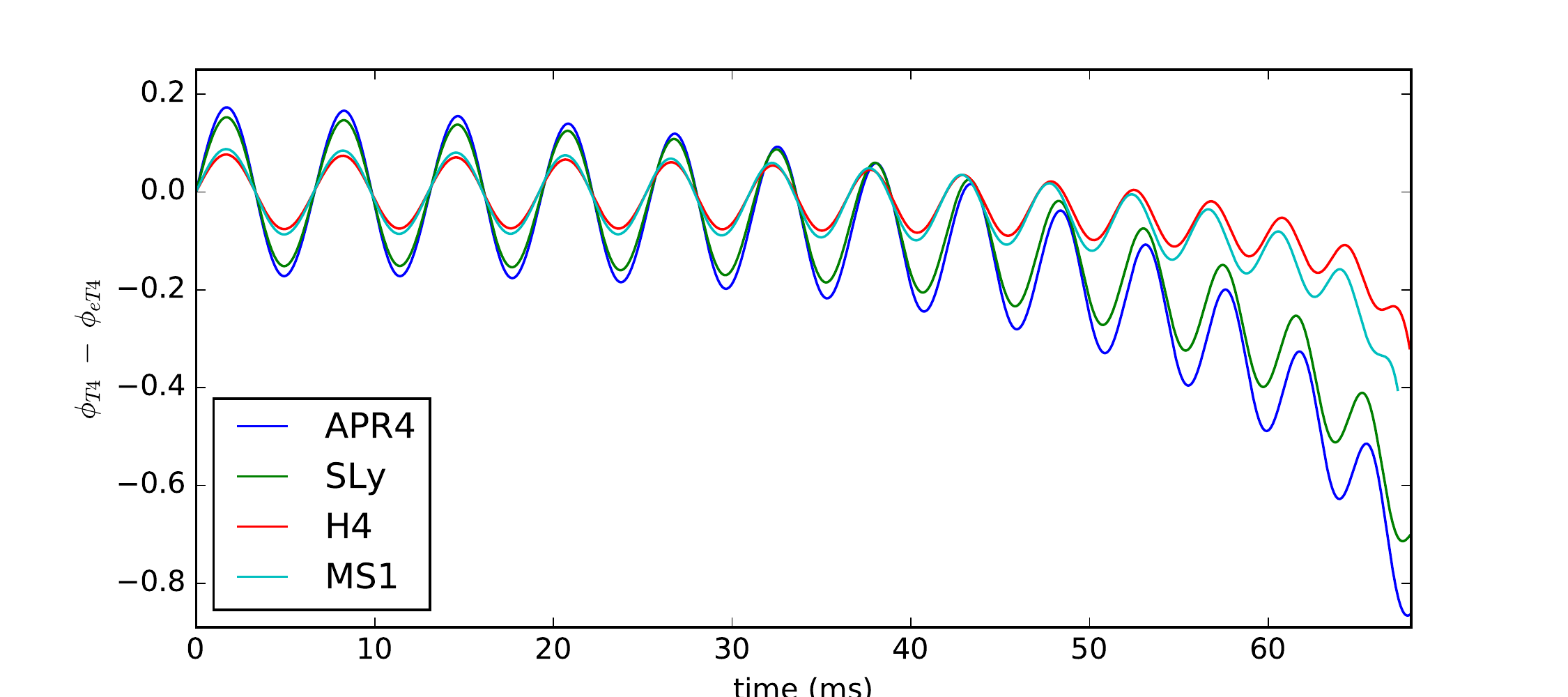}\\
\end{centering}
\vspace{-0.1mm}
\caption{Difference in phase between a eT4 and a standard TaylorT4 approximant at 2PN order (see text). The initial parameters (mass, frequency, eccentricity) are the one for the corresponding simulated model with $d=60$ km.}
\label{fig:AllSimsPN}
\end{figure}

We also compute the difference between the eT4 approximant~\cite{Tanay:2016zog} and a standard circular orbits TaylorT4 waveform~\cite{Boyle:2007ft} at the same Post Newtonian order. Both signals are generated with the \codename{Ligo Algorithms Library} \footnote{The LALSUIT LIGO/Virgo software is publicly available (“git://versions.ligo.org/lalsuite.git”) at the following URL: https://www.lsc-group.phys.uwm.edu/daswg/projects/lalsuite.html.}, considering binaries with the same gravitational mass and initial frequency as our $d=60$ km models, reported in Tab.~\ref{table:InitialData} and \ref{table:EOS}. In Fig.~\ref{fig:AllSimsPN} we plot their difference in phase, starting from the initial frequency of each numerical simulation. For the less compact models with initial eccentricity $~0.01$ (H4 and MS1 EOS) the phase difference oscillates with an amplitude of less than $0.2$ rad, lower than the typical numerical errors on the kind of BNS simulations presented here \cite{Hotokezaka:2015xka,Bernuzzi:2014owa,Hinder:2013oqa}. For the more compact stars (APR4 and SLy EOSs), instead, the phase difference starts to increase in the last orbit, reaching a value around $1$ rad, comparable with the lower bounds on the numerical errors we presented in the right columns of Tab.~\ref{table:errphase} (see next subsection).
Another check was to calculate the detectability of the difference between the circular and the eccentric analytical waveforms, always considering signals starting from the frequencies of our initial data. Even if detectability computations should be done with longer signals, covering all the Advanced Ligo frequency band, to get maximum Signal to Noise ration (SNR), this computation will give us an indication on the importance of eccentricity in currently feasible many-orbits numerical simulations of BNS. This is in particular relevant when planning to use them to calibrate accurate analytical models including also tidal effects \cite{damour:2010effective,Bernuzzi:2014owa,Hinderer:2016eia} or to construct hybrid waveforms for GW detectors data analysis \cite{read:2013matter,Hotokezaka:2016bzh,Radice:2016gym}. We computed the distinguishability of two waveforms $h_1$ and $h_2$ using: 
\begin{equation}
\fl\quad\|h_1 - h_2\|\ =\ \mathrm{min}_{\Delta t,\Delta \phi} \sqrt{4\int_{f_0}^{f_1}{\frac{\left|\tilde{h}_{+,1}(f)-\tilde{h}_{+,2}(f)e^{i\left(2\pi f \Delta t + \Delta \phi\right)}\right|^2}{S_n(f)}\diff f}}.
\end{equation}
Here $\tilde{h}_{+,i}(f)$ is the Fourier transform of the plus polarization of the wave strain, $f_0=9$Hz and $f_1=7000$Hz are the approximate limits for the Advanced Ligo sensitivity band, and $S_n(f)$ is the one sided noise power spectral density for Advanced Ligo in the zero detuning, high power configuration \cite{TheLIGOScientific:2014jea}, often used to check the detectability of effects seen in numerical simulations, for example in~\cite{Takami:2014tva,Hotokezaka:2016bzh}.
It was shown in~\cite{Lindblom:2008cm}, that using this norm, two detected gravitational signals will be distinguishable if $\|h_1 - h_2\| \geq 1$. In particular, in the limit of $\|h_1 - h_2\|=1$, the so-called \textit{marginally distinguishable} case, the two waveforms will be distinguishable with $1\sigma$ statistical significance.

We considered the waveforms produced by an optimally aligned binary at $R=100$Mpc as reference, and we scaled the result to obtain the maximum distance of a binary to be able to marginally detect the effect of the eccentricity values present in our data. Results of this analysis are presented in Tab.~\ref{table:ecc}, where
is evident that the eccentricity effects, although visible by eye in the GW amplitude, would be difficult to detect in the GW interferometers. This is strongly influenced by the fact that current numerical simulations are much shorter than the full evolution of a compact binary system inside all the advanced detectors frequency band. We can conclude that for the spacial resolutions and the number of simulated orbits currently computationally feasible, the errors due to the orbital eccentricity of quasi-circular initial data is less important than the finite-resolution and other numerical errors for the less compact models presented here, with an eccentricity of the order of $0.01$. For the more compact models, instead, with eccentricity around $0.02-0.03$, the phase errors in the last orbit could be important.

All this analysis neglects the interplay between eccentricity and tidal effects, since Post Newtonian approximants including both have not been yet developed.
Nevertheless, eccentricity is a known source of error, and can be reduced with the procedure outlined above~\cite{Dietrich:2015pxa,Kyutoku:2014yba}, which is advisable when calibrating analytical models with long, high-resolution numerical simulations.

\subsection{Comparison of BNS simulations with different starting distance}
\label{sec:comparison}
The main purpose of this work is to compare the dynamical evolution of initial BNS models with the same EOS and different starting distance between the stars (i.e. different initial frequencies). This comparison is useful to get insights on the numerical errors accumulated during many orbits and to validate the correctness of LORENE initial data, even when the two starts are close to each other and the tidal effects have a relevant impact on the system evolution from the beginning of the dynamical simulation. This can also be seen as using the full 3d numerical evolution to fill the gaps between several quasi-equilibrium configurations through which the coalescing binary must pass.

\begin{figure}
\begin{centering}
  \includegraphics[width=\textwidth]{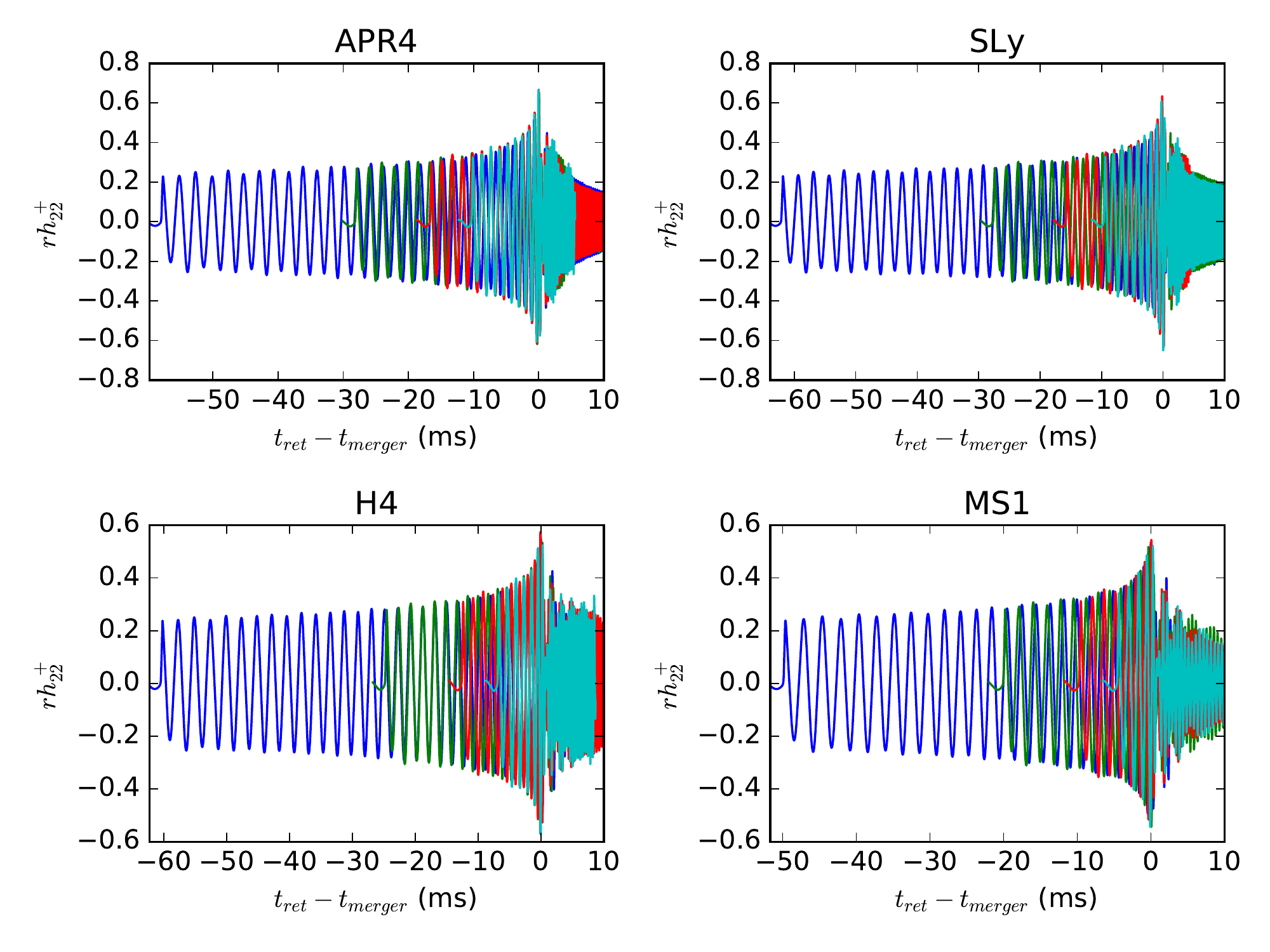}\\
\end{centering}
\vspace{-2mm}
\caption{Overview of the plus polarization of the gravitational wave strain $rh^+_{22}$ for each model simulated. The waveforms of models with the same EOS are aligned at their merger times (the time of the maximum of $\left|h_{22}\right|$). The waves starting from different initial frequencies are marked with different colors (blue: $d=60$ km, green: $d=50$ km, red: $d=44.3$ km, cyan: $d=40$ km). They show a different phase evolution in the last orbits and some differences in the wave amplitude in the merger and post-merger phases.}
\label{fig:AllSims+}
\end{figure}

\begin{figure}
\begin{centering}
  \includegraphics[width=0.95\textwidth]{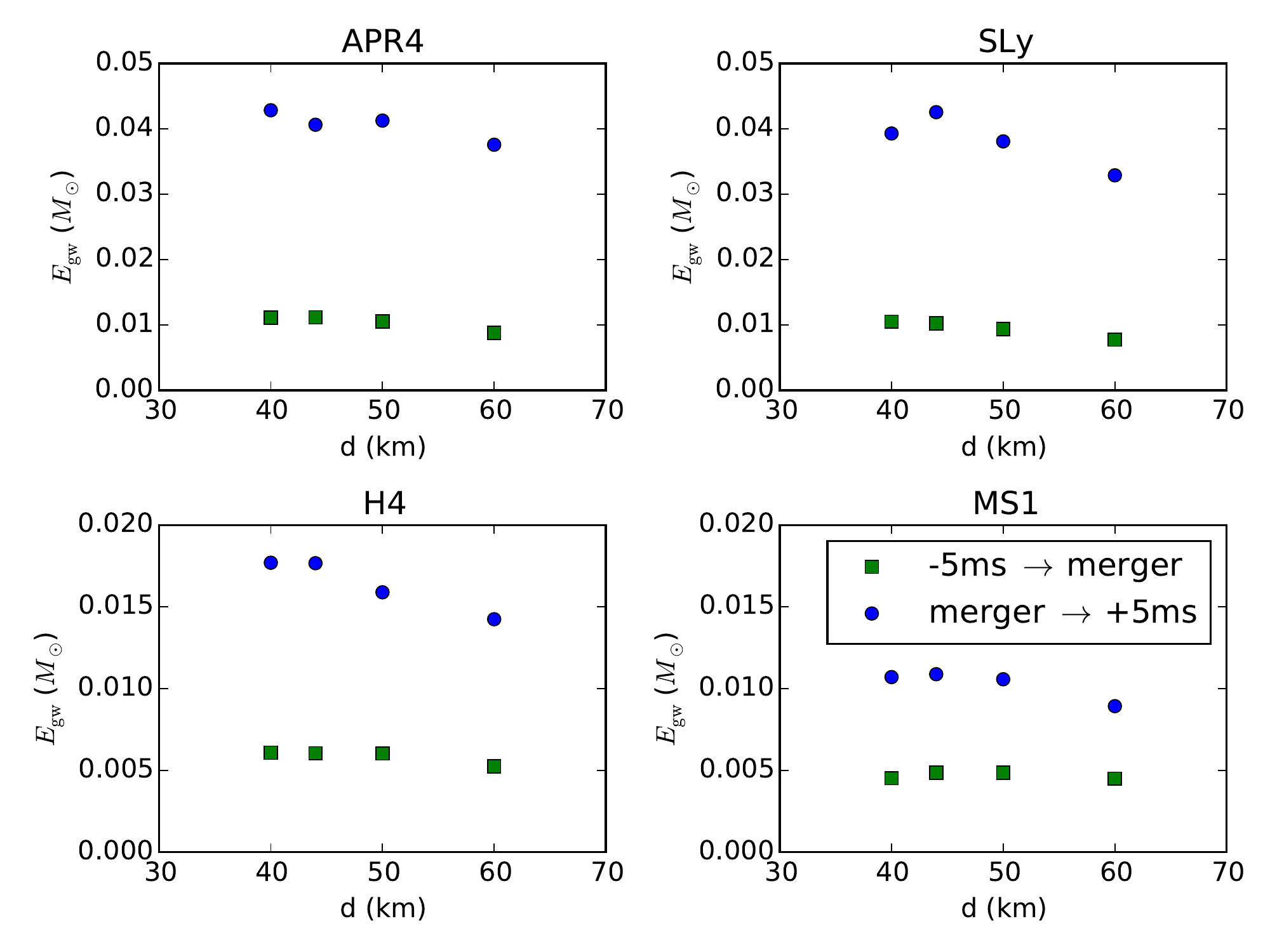}\\
\end{centering}
\vspace{-2mm}
\caption{Radiated energy for all simulated models, computed integrating Eq.~\ref{EQ:dEdt} in time. Green squares represent the energy emitted in the last $5$ms before the merger, while blue dots represent the energy radiated in the first $5$ms after the merger.}
\label{fig:AllSimsEnergy}
\end{figure}

Figure \ref{fig:AllSims+} offers a first global overview of all the simulations we preformed, showing the evolution of the gravitational wave strain polarization $h+$. It is quite obvious from the figure that all simulations with the same EOS agree well during the first part of the coalescence, and start to diverge in the plunge phase, where simulations starting from closer interbinary distances have a slower phase evolution. It is also noticeable that, contrary to what could be expected at first, the signal amplitude after the merger is influenced by the starting frequency. We have quantitatively evaluated this effect (see Fig.~\ref{fig:AllSimsEnergy}), which could turn out to be important in studies investigating the merger remnant energy balance, like~\cite{Bernuzzi:2015opx,Palenzuela:2015dqa,Foucart:2015gaa,Sekiguchi:2015dma,Sekiguchi:2016bjd,Radice:2016gym}.

In order to further investigate the differences in the phase evolution, we aligned the waves from simulations starting at $d=50$ km, $d=44.3$ km and $d=40$ km with the ones starting at $d=60$ km, using the same EOS. First, we considered the gravitational wave strain with respect to the variable \beq{\tilde{t} = t_\mathrm{ret}-t_\mathrm{merger}(d),} namely, aligning the waveforms at the time of merger, as shown in Fig.~\ref{fig:AllSims+}. We define $t_\mathrm{merger}$ as the time for which the amplitude of the GW strain is largest. Then, following a standard procedure often used in the literature~\cite{Boyle:2008ge,Bernuzzi:2012ci,Baiotti:2010xh}, we found the time shift $\Delta t$ and the phase shift $\Delta\phi$, after defining
\begin{equation}
\phi_\mathrm{al_1}(t) = \phi_d(t-\Delta t)+\Delta\phi - \phi_{d=60}(t)
\label{eq:al_shift}
\end{equation}
and then minimizing the following integral:
\begin{equation}
I(\Delta t,\Delta\phi)\ =\ \int_{t_1}^{t_2}{\left|\phi_\mathrm{al_1}(\tilde t)\right|^2}\diff\tilde t
\label{eq:shift}
\end{equation}
between $t_1=3$ms, to avoid the initial spurious radiation, and $t_2=\min(20\mathrm{ms},t_\mathrm{merger}-2\mathrm{ms})$, to avoid the plunge phase.

We computed the difference between the aligned waves phases and the phase of the $d=60$ km simulation. 
\begin{figure}
\begin{centering}
  \includegraphics[width=\textwidth]{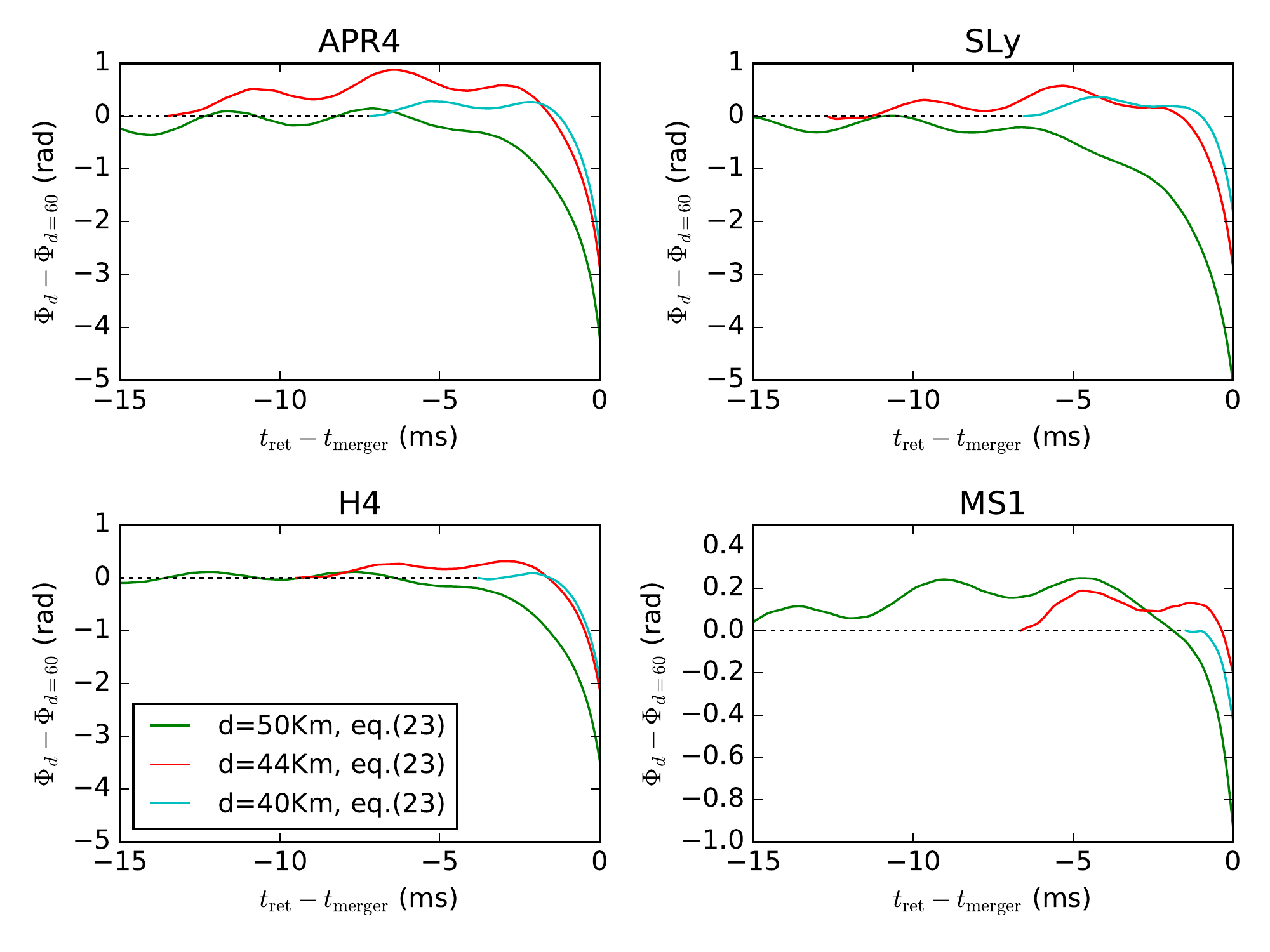}\\
\end{centering}
\vspace{-2mm}
\caption{Phase differences between models with initial interbinary separations of $d=40,44.3,50$ km and the one with initial separation $d=60$ km with the same EOS. They are computed aligning the waveforms finding the time and phase shifts which minimize eq. \ref{eq:shift}. The calculation of cumulative phases starts at the beginning of the alignment interval $t_\mathrm{ret}=t_1=3$ms, in order to not include the initial spurious radiation in the comparison.}
\label{fig:AllSimsPhase}
\end{figure}
\begin{figure}
\begin{centering}
  \includegraphics[width=\textwidth]{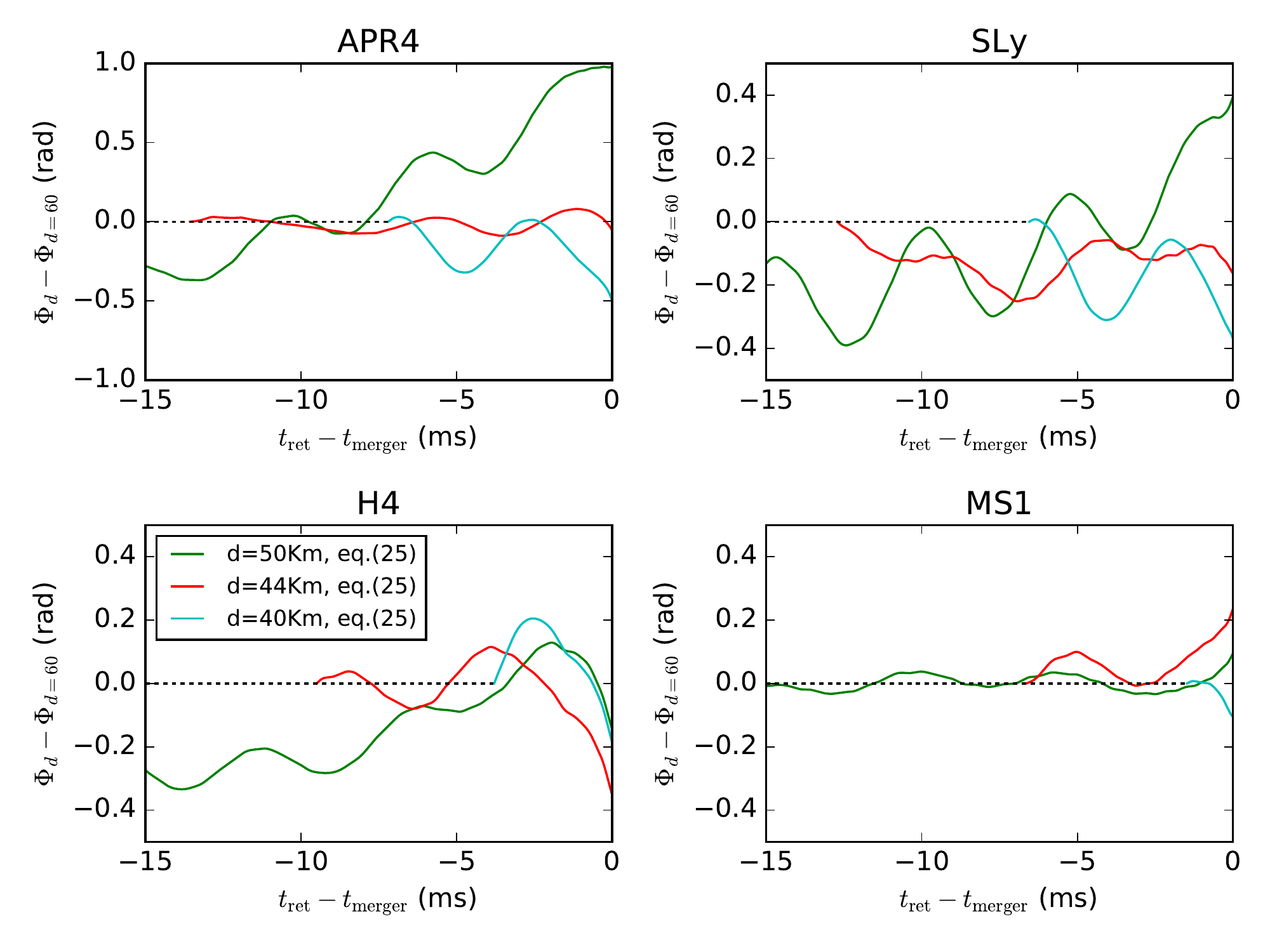}\\
\end{centering}
\vspace{-2mm}
\caption{The same as Fig.~\ref{fig:AllSimsPhase}, but aligning the waveforms finding the time dilatation and the phase shift which minimize Eq.~\ref{eq:dilat}. The resulting phase differences between simulations with different initial frequency can be seen as a lower bound on finite-resolution numerical errors.}
\label{fig:AllSimsPhase_dilat}
\end{figure}
The results can be seen in Fig.~\ref{fig:AllSimsPhase}. Outside the alignment interval, the phase evolutions are quite different, with maximum differences of the order of some radiant. This effect is much more pronounced for the more compact stars with SLy and APR4 EOSs, which have  a maximum phase error of about $4$ radiant, while the H4 EOS shows an intermediate behavior and the  MS1 EOS has much smaller phase difference depending on the starting frequency, with a maximum phase error less than $1$ rad.

Unsatisfied by these results, we tried a different alignment procedure to underline the physical reasons behind such high phase differences. Instead of allowing for an arbitrary time and phase shift as in Eq. \ref{eq:shift}, we allowed for a dilatation of the time variable, as done in \cite{Hotokezaka:2015xka,Hotokezaka:2016bzh} to align waveforms of simulations with different resolutions, to compensate for their different merger times. For each simulation with $d < 60$ km we found the parameters $\eta$ and $\Delta\phi_2$ which, after defining
\begin{equation}
\phi_\mathrm{al_2}(t) = \phi_d(\eta\cdot t)+\Delta\phi_2 - \phi_{d=60}(t)
\label{eq:al_dilat}
\end{equation}
minimize the following integral:
\begin{equation}
I_2(\eta,\Delta\phi_2)\ =\ \int_{t_1}^{t_2}{\left|\phi_\mathrm{al_2}(\tilde t)\right|^2}\diff\tilde t.
\label{eq:dilat}
\end{equation}

Using these, we show in Fig.~\ref{fig:AllSimsPhase_dilat} the difference of the time-dilatation aligned waveforms and the one starting from $d=60$ km.
This time, the phase differences are much smaller for all the EOSs (compare to Fig.~\ref{fig:AllSimsPhase}), and in general do not show a growth specifically in the plunge phase, outside of the alignment time band. All values of the time dilatation parameters $\eta$ found using Eq.~\ref{eq:dilat} are less than one. These results could be interpreted as the simulations starting from a closer interbinary distance took longer to merge compared to ones starting from further apart, when comparing over the same distance before merger. Intuitively, this can be understood by considering the imperfect initial data. These set the initial radial velocity component to zero, although a perfect quasi-circular inspiral has some non-zero inward-facing radial velocity. The magnitude of this velocity depends on the initial distance, which means that the error from setting this velocity to zero is smaller for larger initial separations. For example, stars in a simulation starting from a distance $d=60$ km will have gained some radial velocity by the time they reach $d=40$ km, and will merge faster than a binary starting with zero radial velocity at this distance.
While this is always true comparing the simulations starting from $d=60$ km with the ones starting closer, the effective merger times of simulations starting from $50$, $44.3$ and $40$ km, instead, do not follow always the same trend, suggesting other sources of error, especially for the smallest initial distances. In reference \cite{Suh:2016ctx} it is also noted that initial BNS data computed in the conformal flatness approximation for
the gravitational potential need to be evolved for more than 3 orbits to reach a true,
stable, quasi-equilibrium configuration. This condition is not (or is barely) satisfied
for our d = 40 km simulations (see table \ref{table:Orbits}).

The minimizing parameters for both alignment procedures, with the effective difference in the merger time caused by the time dilatation alignment, can be found in Tab.~\ref{table:shift}.

\begin{table}
\begin{centering}
\begin{tabular}{|c||cc||ccc|}
\hline
\multirow{2}{*}{Model} & \multicolumn{2}{c||}{Shift alignment (Eq.~\ref{eq:al_shift})} & \multicolumn{3}{c|}{Dilatation alignment (Eq.~\ref{eq:al_dilat})}\\
\cline{2-6}
& $\Delta t$ (ms)& $\Delta\phi$ (rad)& $\eta$ & $\Delta\phi_2$ (rad)& $\Delta t_\mathrm{merger}$ (ms)\\
\hline
APR4(a) & 1.40 & 6.26 & 0.899 & -3.40 & 1.02\\
APR4(b) & 1.71 & 6.07 & 0.912 & -4.81 & 1.45\\
APR4(c) & 1.82 & 5.76 & 0.955 & -3.54 & 1.25\\
\hline
SLy(a) & 1.53 & 6.83 & 0.877 & -3.80 & 1.17\\
SLy(b) & 2.02 & 7.46 & 0.900 & -5.06 & 1.57\\
SLy(c) & 2.34 & 7.51 & 0.947 & -4.16 & 1.46\\
\hline
H4(a) & 2.34 & 7.51 & 0.941 & -1.26 & 0.40\\
H4(b) & 1.56 & 5.82 & 0.930 & -2.67 & 0.87\\
H4(c) & 2.49 & 7.85 & 0.942 & -3.92 & 1.43\\
\hline
MS1(a) & 0.27 & 1.40 & 0.963 & -0.30 & 0.16\\
MS1(b) & 0.24 & 0.85 & 0.983 & -0.51 & 0.16\\
MS1(c) & 0.55 & 1.64 & 0.979 & -1.16 & 0.42\\
\hline
\end{tabular}\\
\caption{The first two columns represent the parameters $\Delta t$ and $\Delta\phi$ which minimizes Eq.~\ref{eq:shift} for each model with $d<60$ km. The next two columns show the parameters $\eta$ and $\Delta\phi_2$ which minimize Eq.~\ref{eq:dilat}. The last column shows the effective merger time difference between the time-dilatation aligned waveform and the original one $\Delta t_\mathrm{merger}=t_\mathrm{merger}(1-\eta)$}
\label{table:shift}
\end{centering}
\end{table}
The remaining phase differences after the time-dilatation alignment, collected in Tab.~\ref{table:errphase}, are due to the finite resolution numerical errors, of which they could be considered a lower bound. As anticipated in Sec.~\ref{sec:extrapolation}, they are much higher than the errors coming from the waveform extraction procedure after the application of the perturbative extrapolation formula, even in the first part of the signal, contrary to what was found in \cite{Chu:2015kft} for binary black hole simulations.

\begin{table}
\begin{centering}
\begin{tabular}{|c|ccc|ccc|}
\hline
\multirow{2}{*}{EOS} &
\multicolumn{3}{c|}{Shift alignment (Eq.~\ref{eq:shift})} &
\multicolumn{3}{c|}{Dilatation alignment (Eq.~\ref{eq:dilat})}\\

& \multicolumn{3}{c|}{$\mathcal{E}_{\phi_\mathrm{al_1}}$(rad)} &
\multicolumn{3}{c|}{$\mathcal{E}_{\phi_\mathrm{al_2}}$(rad)} \\

\cline{2-7}
& $d=50$ km & $44.3$ km & $40$ km & $d=50$ km & $44.3$ km & $40$ km\\

\hline
APR4 &   4.12 & 2.79 & 2.40 & 0.98 & 0.09 & 0.49 \\
\hline
SLy & 5.01 & 2.79 & 1.77 & 0.40 & 0.25 & 0.37 \\
\hline
H4 & 3.41 & 2.06 & 1.85 & 0.33 & 0.35 & 0.20 \\
\hline
MS1 & 0.91 & 0.19 & 0.41 & 0.09 & 0.23 & 0.10\\
\hline
\end{tabular}\\
\end{centering}
\caption{Maximum phase difference in the inspiral phase between waveforms from simulations starting from $d<60$ km and simulations with $d=60$ km, aligned according to Eq.~\ref{eq:shift} (first three columns) or Eq.~\ref{eq:dilat} (second three columns).}
\label{table:errphase}
\end{table}
\begin{figure}
\begin{centering}
  \includegraphics[width=0.95\textwidth]{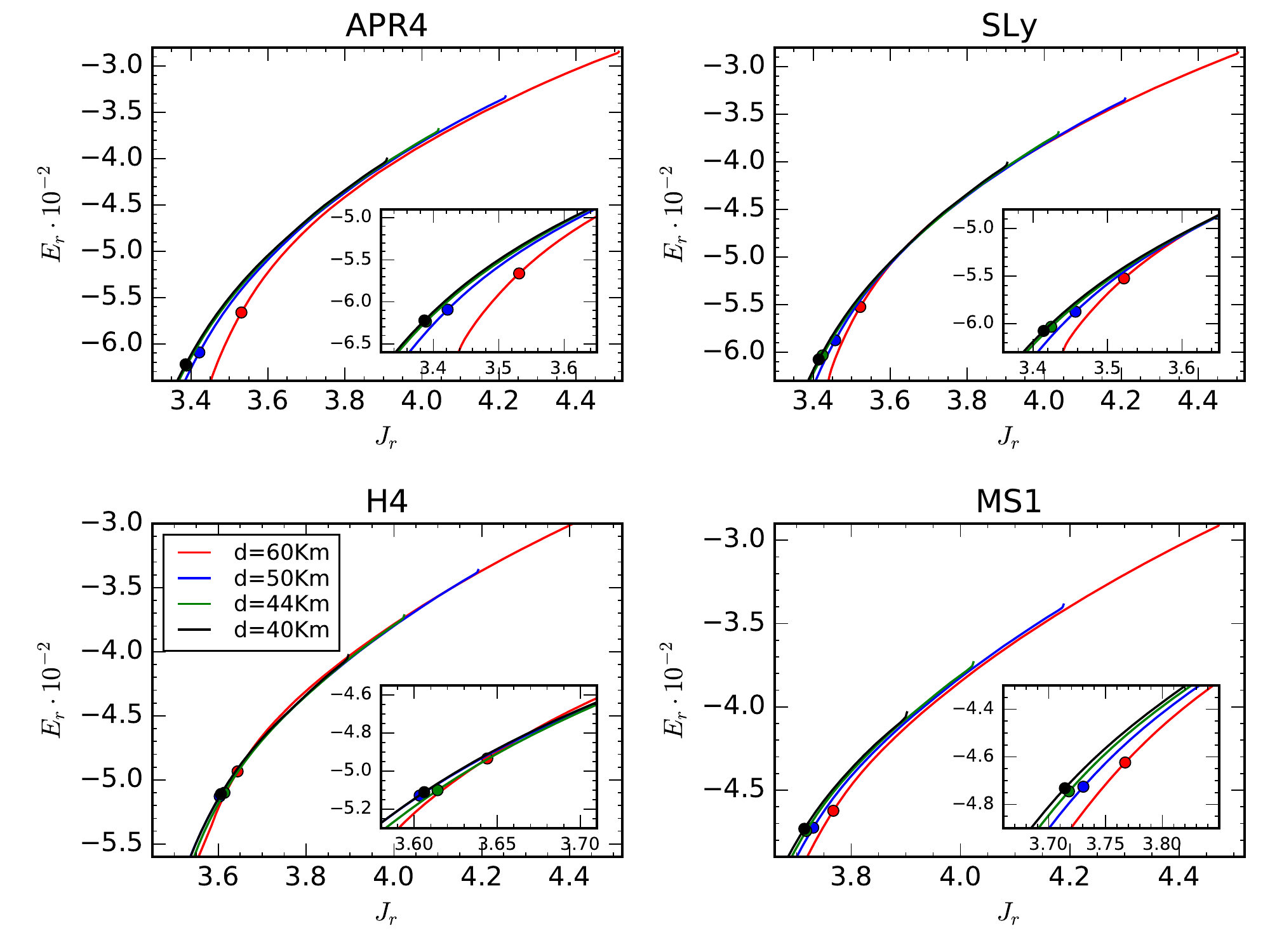}\\
\end{centering}
\vspace{-2mm}
\caption{Reduced adimensional energy vs reduced adimensional angular momentum curves (see text for definitions) for each model simulated. The merger time values are marked with a filled dot the same color of the corresponding curve. All the curves show very good agreement in the inspiral phase and a small departure in the plunge phase.}
\label{fig:AllSimsEJ}
\end{figure}
This picture is consistent with what can be found analyzing the radiated energy and angular momentum, computed from simulations starting at different interbinary separations. In Fig.~\ref{fig:AllSimsEJ} we plotted for each model the gauge invariant relation between the reduced, adimensional binding energy $E_r= \frac{\left(\frac{E}{M}-1\right)}{\nu}$ and the reduced, adimensional angular momentum $J_r=\frac{J}{M^2\nu}$, often used in the literature to compare numerical and analytical waveforms \cite{Damour:2011fu,Bernuzzi:2012ci,Bernuzzi:2014owa}. Here $E = M_\mathrm{ADM}-E_\mathrm{gw}$ and $J = J_\mathrm{ADM}-J_\mathrm{gw}$, where the ADM quantities refer to the initial data and are reported in Tab.~\ref{table:InitialData}, and $E_\mathrm{gw}$, $J_\mathrm{gw}$ are the radiated energy and angular momentum along the rotation axis, calculated integrating Eq.~\ref{EQ:dEdt} in time. $M=M_1+M_2$ is the sum of the gravitational masses of the two stars in isolation, and $\nu$ is the reduced mass divided by $M$: $\nu = \frac{M_1M_2}{M^2}$.

For all EOSs, $E_r(J_r)$ of the simulations with different starting frequency agrees very well, at least until the start of the simulation with $d=40$ km. After that point the simulations with $d=60$ km start to diverge, resulting in a merger at a higher energy and angular momentum. The other three simulations, instead, agree much better until the merger, and their merger energies and angular momenta are also close to each other. This has the natural interpretation that the simulations with a longer effective merger time emit more energy and angular momentum during their longer and slower approach to the merger.

In all the simulations starting from $d<60$ km we observe a good agreement of the initial energy and angular momentum, after the short relaxation period, with the simulations starting from a larger interbinary distance, with a maximum difference of a few tenths of percent. This means that, at least in the inspiral phase, our numerical evolutions, even at this modest resolution, are able to join the sequence of quasi-equilibrium states generated by the LORENE library. However, in the late inspiral and plunge phase, when the tidal effects start to be relevant, there are important differences. In principle, this could be due to higher finite-resolution numerical errors accumulated by the longer simulations and we would like to note that in our previous work~\cite{DePietri:2015lya} we showed that, with this combination of numerical methods, the merger time of a simulation increases at better resolutions. However
an another possible contribution or explanation to differences between simulations starting from different interbinary distances may come from tidal effects that might not be properly developed and resolved in simulations starting from higher frequencies (or which is the same starting from closer distance). Tidal effects are already important at the start of the $d=40$ km simulations. On the other hand, they could be under-resolved in the calculation of initial data, which uses an approximate conformal flatness treatment of the gravitational potential. As we have shown before, the more compact stars are the ones for which we see higher deviations in the $d=60$ km simulations with respect to the ones starting from further inward.

The role of dynamical tidal effects was recently also underlined in~\cite{Hinderer:2016eia}. These effects develop from the interaction of the tidal field with the star's quasi-normal modes of oscillation. These dynamic tides could be developed during a many orbits evolution, but are not considered in the calculation of the initial data used here. However, they should be more important for the less compact stars. Only further investigations with higher resolutions and different EOSs can give a definitive answer on the physical reasons behind the differences in numerical simulations of BNS starting with different frequencies and on the cheapest acceptable initial interbinary distance to perform accurate studies of the merger and post-merger phases.

In particular, new simulations with more accurate numerical methods could help reducing the numerical errors accumulated during the longer simulations with $d>50$ km, making them  reliable also at the modest resolutions employed here, if that turns out to be the biggest source of error. Possible improvements over the numerical methods employed in this work are the use of the Z4 family of formulations for the Einstein Equations \cite{Alic:2011gg,Weyhausen:2011cg} (due to their constraint violations damping capabilities), and the use of higher-than-second order convergent schemes for hydrodynamics, as introduced in ~\cite{Radice:2012cu}.
These studies are beyond our current computational capabilities, but we plan to implement them in future works.

Finally, we analyzed the effect of the different initial interbinary distance on the post-merger gravitational signal. We computed the energy radiated by the different simulations in different time intervals around the merger. The results are plotted in Fig.~\ref{fig:AllSimsEnergy}. We see for all EOSs that even in the first part of the post-merger phase, where most of the gravitational wave energy is emitted according to~\cite{Bernuzzi:2015opx}, there are big differences in the energy radiated by simulations starting with different initial frequencies. In particular, all simulations with $d=60$ km emit less energy than the others. These differences in the immediate post-merger phase are higher than the ones in the late inspiral phase. This could have an important impact in the studies on the energy balance in the post-merger phase~\cite{Bernuzzi:2015opx,Palenzuela:2015dqa,Foucart:2015gaa,Sekiguchi:2015dma,Sekiguchi:2016bjd,Radice:2016gym}.

In addition, new simulations involving more microphysical ingredients such as finite-temperature nuclear EOSs, neutrino emission and magnetic fields are needed to quantify the impact of the initial orbital frequency on the post-merger physical observables.

\begin{figure}
\begin{centering}
  \includegraphics[width=\textwidth]{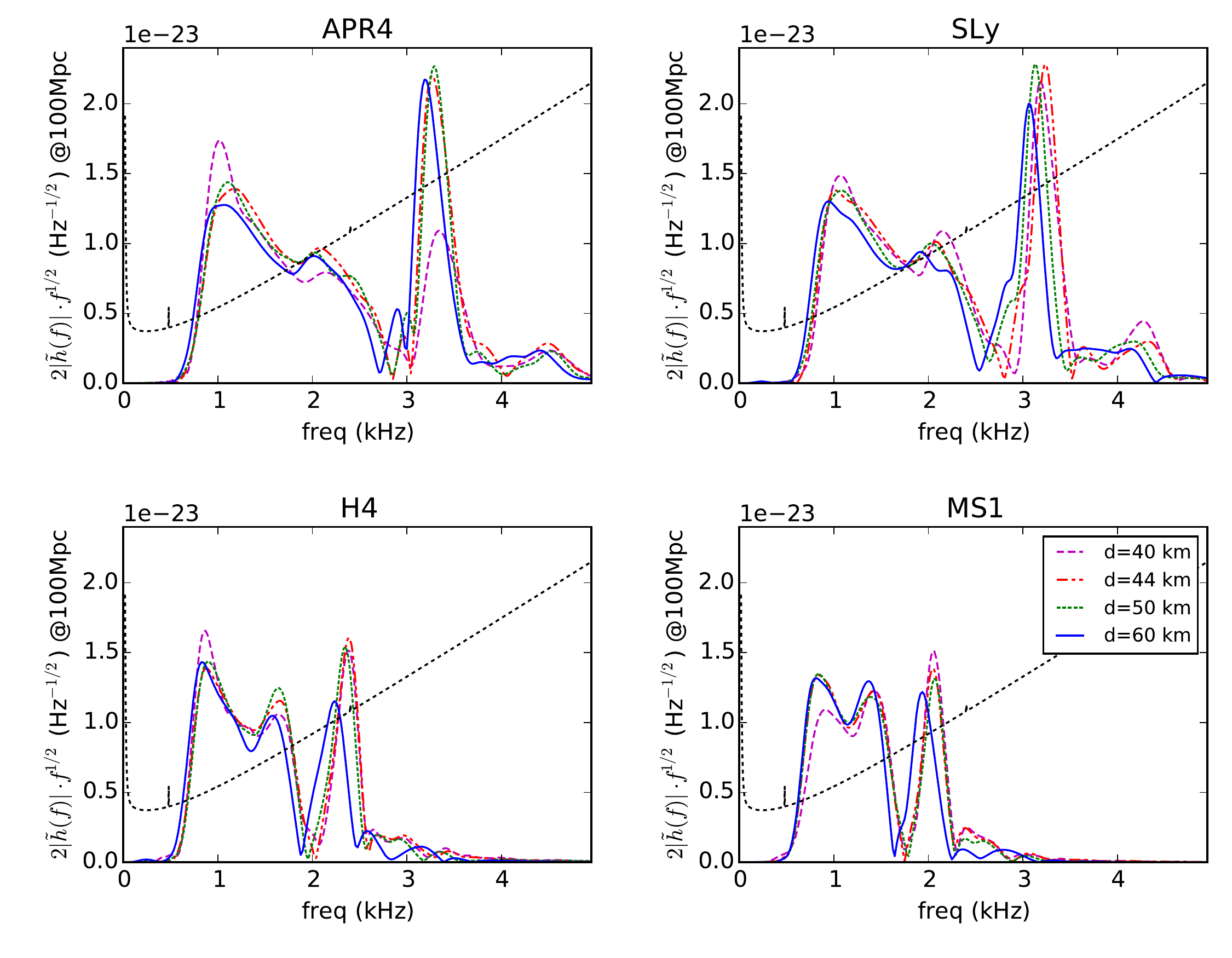}\\
\end{centering}
\vspace{-2mm}
\caption{Amplitude spectral densities of the gravitational wave signal from $8$ms before the merger to $8$ms after it, for a source at $100$Mpc. For some models the total signal power is lower either because the simulations were shorter in the post-merger phase (H4(d),APR4(a)), or because their merger time was less than $8$ms (H4(a),MS1(a)). The black dashed curve represents the sensibility of Advanced Ligo in the Zero detuning-high power configuration \cite{TheLIGOScientific:2014jea}.}
\label{fig:psd}
\end{figure}
We also analyzed the spectrum of the gravitational waves emitted in the post-merger phase by the deformed, differentially rotating HMNS. EOS effects were already reported in studies about single neutron stars bar-deformed due to dynamical instabilities \cite{DePietri:2014mea,Loffler:2014jma}.        
In several recent works~\cite{Takami:2014tva,Rezzolla:2016nxn,kastaun:2015properties,bauswein:2015unified,bernuzzi:2015modeling,read:2013matter}, the spectral peaks of the post-merger BNS gravitational signal were correlated with stellar characteristics linked with the EOS (compactness, tidal deformability). In Fig.~\ref{fig:psd} we show the amplitude spectral density of the gravitational waves from our simulations $\left|\tilde{h}\right|f^{1/2}$, where
\begin{equation}
\left|\tilde{h}\right|\ =\ \sqrt{\frac{\left|\tilde{h}_+\right|^2\ +\ \left|\tilde{h}_{\times}\right|^2}{2}}.
\end{equation}
We compute it in the interval between $8$ms before the merger and $8$ms after it, applying first a Blackman window function to the signal. All simulations show a dominant peak, corresponding to the fundamental $m=2$ oscillation mode of the bar-deformed rotating neutron star created after the merger.

Models with more compacts stars (APR4, SLy EOSs) have the dominant peak ($f_{peak}$) at higher frequencies and show two, less clearly identifiable, secondary peaks ($f_-$ and $f_+$) to both sides of $f_{peak}$, almost equidistant. The subdominant peaks, in particular the one at $f_+$, are broad and difficult to identify because of a low resolution of our Fourier transform ($62$Hz for models covering the $\left[t_m-8,t_m+8\right]$ interval), since in this work we concentrated more on the inspiral dynamics and didn't simulate for a long time the post-merger evolution.
Less compact stars (H4 and MS1 EOSs), instead, show only one clear, secondary peak at lower frequency than $f_{peak}$, with no apparent corresponding peak at higher frequency.

There is a debate in the recent literature about the physical origin of these secondary peaks. They are interpreted as being generated by the modulation of the oscillation of the two rotating stellar cores~\cite{Takami:2014tva,Rezzolla:2016nxn,kastaun:2015properties}, or as coupling between the fundamental $m=2$ mode and the quasiradial $m=0$ mode~\cite{Stergioulas:2011gd}, or, in the low frequency peak case, for the less compact stars, as due to the faster rotation of the spiral arms with respect to the inner double-core structure~\cite{bauswein:2015unified}. Our simulations are too short to be able to differentiate between these proposals, which is made more difficult by the fact that the theoretical values for the frequencies of these proposed effects are often close~\cite{Rezzolla:2016nxn}.

We verified, however, that the frequencies of the peaks for our $d=44.3$ km simulations remarkably agree (within our frequency errors) with the ones reported in~\cite{Takami:2014tva,Rezzolla:2016nxn} (where a much higher resolution is used), for their models {APR4-q10-M1275}, 
{SLy-q10-M1275}, {H4-q10-M1300}, which start from a interbinary distance of $d=45$ km.

Focusing on the dominant peak, we noticed that its frequency has a dependence on the initial simulation frequency, with differences between simulations with the same EOS higher than our Fourier transform uncertainties. We list these results in Tab.~\ref{table:fft}. Models starting from $d=60$ km have a consistently lower frequency of the dominant spectral peak than models starting closer together. This is linked with their lower instantaneous frequency at the time of the merger, when the deformed hyper-massive neutron star is formed. These differences could also be related to the higher accumulated numerical errors in the longer simulations, as hypothesized before to explain the phase differences in the inspiral part of the signal. This post-merger effect should be checked again in simulations with higher resolutions and post-merger simulated time, to measure $f_{peak}$ in the quasi-stationary phase of the HMNS evolution, after the relaxation of all the other modes. That is, however, outside the scope of this work, concentrating on the coalescent phase.
\begin{table}
\begin{centering}
\begin{tabular}{|c|c|c|c|}
\hline
Model & $f_p$ (kHz) & $f_{-}$ (kHz) & $f_\mathrm{merger}$ (kHz)  \\
\hline
APR4(a) &   3.34  &  2.14  &  1.97   \\
APR4(b) &   3.26  &  2.08  &  1.97  \\
APR4(c) &   3.29  &  2.02  &  1.94  \\
APR4(d) &   3.19  &  2.02  &  1.87  \\
\hline
SLy(a) &  3.19   &  2.15 &  1.92  \\
SLy(b) &  3.24  &   2.08 &  1.88  \\
SLy(c) &  3.13   &  2.03 &  1.87   \\
SLy(d) &  3.07   &  1.92 &  1.72  \\
\hline
H4(a) &   2.38  &  1.64  &  1.44  \\
H4(b) &  2.39   &  1.66  &  1.46   \\
H4(c) &   2.35  &  1.64  &  1.44  \\
H4(d) &   2.24  &  1.58  &  1.41  \\
\hline
MS1(a) &  2.05   &  1.42  & 1.29  \\
MS1(b) &   2.05  &  1.43  & 1.33  \\
MS1(c) &   2.07  &  1.39  & 1.32   \\
MS1(d) &   1.93  &  1.37  & 1.26  \\
\hline
\end{tabular}\\
\end{centering}
\caption{Results from the spectral analysis. In the first column we show the frequency of the dominant post-merger peak, calculated taking the frequency of the maximum of the amplitude spectral density after interpolation with a cubic spline. The second column shows the frequency of the left secondary peak $f_-$ (the $f_1$ peak for \cite{Takami:2014tva,Rezzolla:2016nxn}). In the third column we report the instantaneous frequency at merger, computed taking the derivative of the phase evolution. The uncertainty of the Fourier transform frequencies are in the interval $[62,75]$ Hz. Some models have a higher frequency uncertainty because they either have shorter post-merger simulations (APR4(a),H4(d)) or they have a merger time less than $8$ms (H4(a),MS1(a)). }
\label{table:fft}
\end{table}

\section{Conclusions}
We presented results from numerical simulations of BNS mergers using
four different nuclear EOSs for the NS interior and starting from four
different values for the interbinary distance.

We tested different procedures for the extrapolation
of the gravitational wave signal $\psi_4$, extracted at finite
radii, to null infinity. In particular, we used for the first time in BNS simulations
the second order perturbative extrapolation of~\cite{Nakano:2015perturbative}, show-casing advantages in
reducing the extraction errors with respect to the first-order extrapolation~\cite{Lousto:2010qx}.
We also showed that the error of the
wave extraction procedure, after the extrapolation, is smaller than the typical numerical errors
reported for this kind of BNS simulations, even
in the first part of the signal.

We proposed a simple procedure to integrate the Newman-Penrose scalar
$\psi_4$ to obtain the gravitational wave strain $h$, based on
applying a high-pass digital filter to $h$ after a time-domain
integration, instead of filtering $\psi_4$ before a frequency domain
integration, which currently is the most common procedure in numerical
relativity \cite{Reisswig:2011notes}. Our procedure gives the cleanest
signal when applying it after the extrapolation formula, whose integral
terms introduce additional low-frequency noise.

We computed the small orbital eccentricity due to the
quasi-circular initial data for all models, showing a dependence on
the NS EOS but not being able to show any clear dependence on the
starting distance between the stars. We compared the amplitude of our
simulations with a recently developed analytical model for eccentric
binaries \cite{Tanay:2016zog}, showing good agreement in the first
part of the signal. This is a confirmation of the fact that the
gravitational wave amplitude oscillations are really an imprint of
the orbital eccentricity and of the quality of the simple,
approximate, eccentricity calculation procedure we used. This result
confirms also that initial data eccentricity is, until the very last
few orbits, the main source of discrepancy between gravitational waves
computed with numerical relativity and point-particle Post-Newtonian
approximations.

In order to understand the relevance of this measured initial data eccentricity, 
we compared two analytical waveforms at the same PN order, with the
same mass and initial frequency of our $d=60$ km models, one with the
same eccentricity of our numerical simulations. Their difference in phase
is almost always below typical numerical errors in current BNS simulations,
except for the last orbit of models with the same eccentricity of our more
compact ones (with APR4 and SLy EOSs). The
detectability of their difference in Advanced Ligo is small, requiring
a maximum distance of $33-67$ Mpc for its detection at $1\sigma$. This
is due to the shortness of current numerical simulations with respect to
the whole frequency band of interferometric detectors. The effect of
eccentricity will be more important when longer numerical simulations
will be performed for GW detector
data analysis. Techniques for reducing initial data eccentricity have been
developed by some groups in the last couple of years
\cite{Dietrich:2015pxa,Kyutoku:2014yba}. Since the recent beginning of
the advanced detector era, we see 
an urgent need for public low-eccentricity BNS initial data
in the numerical relativity community.

We compared the results of simulations with the same NS EOS
and different starting interbinary distance. Their emitted
gravitational signals show large phase differences in the last
orbits before merger, when aligning them with the standard formula,
allowing only for a relative shift in the time
variable. When we aligned them allowing
for a time dilatation, to account for the longer effective merger time
of simulations with $d<60$ km, the remaining phase differences were
close to the expected numerical errors. This effective merger time
difference was found to be higher for more compact stars.

Investigating the post-merger spectrum, we found that simulations starting 
from $d=60$ km have the dominant peak at a somewhat lower frequency 
(but still consistent with the error) than the others, while we found a
very good agreement with the results of~\cite{Takami:2014tva,Rezzolla:2016nxn} 
for simulations starting from a distance of $d=44.3$ km.

Our results also suggest that additional effects can be present 
in BNS simulations with more than 10 orbits, but an investigation on these
is left for future work.

\label{sec:conclusion}

\ack

This project greatly benefited from the availability of public software that
enabled us to conduct all simulations, namely ``LORENE'' and the ``Einstein
Toolkit''. We express our gratitude the many people that contributed to
their realization.

This work would have not been possible without the support of the SUMA INFN
project that provided the financial support of the work of AF and the computer
resources of the CINECA ``GALILEO'' HPC Machine, where part of the simulations
were performed. Other computational resources were provided by he Louisiana
Optical Network Initiative (QB2, allocations loni\_hyrel15, and
loni\_hyrel16), and by the LSU HPC facilities (SuperMike II, allocations
hpc\_hyrel15 and hpc\_hyrel16). FL is directly supported by, and this project
heavily used infrastructure developed using support from the National Science
Foundation in the USA (Grants No. 1212401, No. 1212426, No. 1212433, No.
1212460). Partial support from INFN ``Iniziativa Specifica TEONGRAV'' and by
the ``NewCompStar'', COST Action MP1304, are kindly acknowledged.

\providecommand{\newblock}{}


\begin{thebibliography}{100}
\expandafter\ifx\csname url\endcsname\relax
  \def\url#1{{\tt #1}}\fi
\expandafter\ifx\csname urlprefix\endcsname\relax\def\urlprefix{URL }\fi
\providecommand{\eprint}[2][]{\url{#2}}

\bibitem{Abbott:2016blz}
Abbott B~P {\em et~al.\/} (Virgo, LIGO Scientific) 2016 {\em Phys. Rev.
  Lett.\/} {\bf 116} 061102 (\textit{Preprint} \eprint{1602.03837})

\bibitem{TheLIGOScientific:2014jea}
Aasi J {\em et~al.\/} (LIGO Scientific) 2015 {\em Class. Quant. Grav.\/} {\bf
  32} 074001 (\textit{Preprint} \eprint{1411.4547})

\bibitem{TheVirgo:2014hva}
Acernese F {\em et~al.\/} (VIRGO) 2015 {\em Class. Quant. Grav.\/} {\bf 32}
  024001 (\textit{Preprint} \eprint{1408.3978})

\bibitem{LIGOVIRGO:2013}
{LIGO Scientific Collaboration}, {Virgo Collaboration}, {Aasi} J, {Abadie} J,
  {Abbott} B~P, {Abbott} R, {Abbott} T~D, {Abernathy} M, {Accadia} T,
  {Acernese} F and et~al 2013 {\em ArXiv e-prints\/} (\textit{Preprint}
  \eprint{1304.0670})

\bibitem{Shibata:1999wm}
Shibata M and Uryu K 2000 {\em Phys. Rev. D\/} {\bf 61} 064001
  (\textit{Preprint} \eprint{arXiv:gr-qc/9911058})

\bibitem{Pretorius:2005gq}
Pretorius F 2005 {\em Phys. Rev. Lett.\/} {\bf 95} 121101 (\textit{Preprint}
  \eprint{arXiv:gr-qc/0507014})

\bibitem{Campanelli:2005dd}
Campanelli M, Lousto C~O, Marronetti P and Zlochower Y 2006 {\em Phys. Rev.
  Lett.\/} {\bf 96} 111101 (\textit{Preprint} \eprint{arXiv:gr-qc/0511048})

\bibitem{Baker:2005vv}
Baker J~G, Centrella J, Choi D~I, Koppitz M and van Meter J 2006 {\em Phys.
  Rev. Lett.\/} {\bf 96} 111102 (\textit{Preprint}
  \eprint{arXiv:gr-qc/0511103})

\bibitem{EinsteinToolkit:web}
{Einstein Toolkit}: Open software for relativistic astrophysics
  \urlprefix\url{http://einsteintoolkit.org/}

\bibitem{Loffler:2011ay}
L{\"{o}}ffler F, Faber J, Bentivegna E, Bode T, Diener P, Haas R, Hinder I,
  Mundim B~C, Ott C~D, Schnetter E, Allen G, Campanelli M and Laguna P 2012
  {\em Class. Quantum Grav.\/} {\bf 29} 115001 (\textit{Preprint}
  \eprint{arXiv:1111.3344 [gr-qc]})

\bibitem{Moesta:2013dna}
M{\"o}sta P, Mundim B~C, Faber J~A, Haas R, Noble S~C, Bode T, L{\"o}ffler F,
  Ott C~D, Reisswig C and Schnetter E 2014 {\em Classical and Quantum
  Gravity\/} {\bf 31} 015005 (\textit{Preprint} \eprint{arXiv:1304.5544
  [gr-qc]})

\bibitem{lorene:web}
{LORENE}: {L}angage {O}bjet pour la {RE}lativit\'e {N}um\'eriqu{E}
  \urlprefix\url{http://www.lorene.obspm.fr/}

\bibitem{Gourgoulhon:2000nn}
Gourgoulhon E, Grandclement P, Taniguchi K, Marck J~A and Bonazzola S 2001 {\em
  Phys. Rev. D\/} {\bf 63} 064029 (\textit{Preprint}
  \eprint{arXiv:gr-qc/0007028})

\bibitem{DePietri:2015lya}
De~Pietri R, Feo A, Maione F and Löffler F 2016 {\em Phys. Rev.\/} {\bf D93}
  064047 (\textit{Preprint} \eprint{1509.08804})

\bibitem{read:2013matter}
Read J~S, Baiotti L, Creighton J~D~E, Friedman J~L, Giacomazzo B, Kyutoku K,
  Markakis C, Rezzolla L, Shibata M and Taniguchi K 2013 {\em Phys. Rev.\/}
  {\bf D88} 044042 (\textit{Preprint} \eprint{1306.4065})

\bibitem{Hotokezaka:2013mm}
Hotokezaka K, Kyutoku K and Shibata M 2013 {\em Phys. Rev.\/} {\bf D87} 044001
  (\textit{Preprint} \eprint{1301.3555})

\bibitem{Hotokezaka:2015xka}
Hotokezaka K, Kyutoku K, Okawa H and Shibata M 2015 {\em Phys. Rev.\/} {\bf
  D91} 064060 (\textit{Preprint} \eprint{1502.03457})

\bibitem{Baiotti:2011am}
Baiotti L, Damour T, Giacomazzo B, Nagar A and Rezzolla L 2011 {\em Phys. Rev.
  D\/} {\bf 84} 024017 (\textit{Preprint} \eprint{arXiv:1103.3874 [gr-qc]})

\bibitem{bauswein:2015unified}
Bauswein A and Stergioulas N 2015 {\em Phys. Rev.\/} {\bf D91} 124056
  (\textit{Preprint} \eprint{1502.03176})

\bibitem{Bauswein:2015vxa}
Bauswein A, Stergioulas N and Janka H~T 2016 {\em Eur. Phys. J.\/} {\bf A52} 56
  (\textit{Preprint} \eprint{1508.05493})

\bibitem{Takami:2014tva}
Takami K, Rezzolla L and Baiotti L 2015 {\em Phys. Rev.\/} {\bf D91} 064001
  (\textit{Preprint} \eprint{1412.3240})

\bibitem{kastaun:2015properties}
Kastaun W and Galeazzi F 2015 {\em Phys. Rev.\/} {\bf D91} 064027
  (\textit{Preprint} \eprint{1411.7975})

\bibitem{bauswein:2014revealing}
Bauswein A, Stergioulas N and Janka H~T 2014 {\em Phys. Rev.\/} {\bf D90}
  023002 (\textit{Preprint} \eprint{1403.5301})

\bibitem{hotokezaka:2013remnant}
Hotokezaka K, Kiuchi K, Kyutoku K, Muranushi T, Sekiguchi Y~i, Shibata M and
  Taniguchi K 2013 {\em Phys. Rev.\/} {\bf D88} 044026 (\textit{Preprint}
  \eprint{1307.5888})

\bibitem{Lehner:2016lxy}
Lehner L, Liebling S~L, Palenzuela C, Caballero O~L, O'Connor E, Anderson M and
  Neilsen D 2016  (\textit{Preprint} \eprint{1603.00501})

\bibitem{Foucart:2015gaa}
Foucart F, Haas R, Duez M~D, O’Connor E, Ott C~D, Roberts L, Kidder L~E,
  Lippuner J, Pfeiffer H~P and Scheel M~A 2016 {\em Phys. Rev.\/} {\bf D93}
  044019 (\textit{Preprint} \eprint{1510.06398})

\bibitem{Bernuzzi:2015opx}
Bernuzzi S, Radice D, Ott C~D, Roberts L~F, Moesta P and Galeazzi F 2015
  (\textit{Preprint} \eprint{1512.06397})

\bibitem{Palenzuela:2015dqa}
Palenzuela C, Liebling S~L, Neilsen D, Lehner L, Caballero O~L, O'Connor E and
  Anderson M 2015 {\em Phys. Rev.\/} {\bf D92} 044045 (\textit{Preprint}
  \eprint{1505.01607})

\bibitem{Sekiguchi:2015dma}
Sekiguchi Y, Kiuchi K, Kyutoku K and Shibata M 2015 {\em Phys. Rev.\/} {\bf
  D91} 064059 (\textit{Preprint} \eprint{1502.06660})

\bibitem{Sekiguchi:2016bjd}
Sekiguchi Y, Kiuchi K, Kyutoku K, Shibata M and Taniguchi K 2016
  (\textit{Preprint} \eprint{1603.01918})

\bibitem{Kiuchi:2015sga}
Kiuchi K, Cerdá-Durán P, Kyutoku K, Sekiguchi Y and Shibata M 2015 {\em Phys.
  Rev.\/} {\bf D92} 124034 (\textit{Preprint} \eprint{1509.09205})

\bibitem{Dionysopoulou:2015tda}
Dionysopoulou K, Alic D and Rezzolla L 2015 {\em Phys. Rev.\/} {\bf D92} 084064
  (\textit{Preprint} \eprint{1502.02021})

\bibitem{Ponce:2014sza}
Ponce M, Palenzuela C, Lehner L and Liebling S~L 2014 {\em Phys. Rev.\/} {\bf
  D90} 044007 (\textit{Preprint} \eprint{1404.0692})

\bibitem{Giacomazzo:2014qba}
Giacomazzo B, Zrake J, Duffell P, MacFadyen A~I and Perna R 2015 {\em
  Astrophys. J.\/} {\bf 809} 39 (\textit{Preprint} \eprint{1410.0013})

\bibitem{Kiuchi:2015qua}
Kiuchi K, Sekiguchi Y, Kyutoku K, Shibata M, Taniguchi K and Wada T 2015 {\em
  Phys. Rev.\/} {\bf D92} 064034 (\textit{Preprint} \eprint{1506.06811})

\bibitem{Palenzuela:2013hu}
Palenzuela C, Lehner L, Ponce M, Liebling S~L, Anderson M, Neilsen D and Motl P
  2013 {\em Phys. Rev. Lett.\/} {\bf 111} 061105 (\textit{Preprint}
  \eprint{1301.7074})

\bibitem{Rezzolla:2011da}
Rezzolla L, Giacomazzo B, Baiotti L, Granot J, Kouveliotou C and Aloy M~A 2011
  {\em Astrophys. J.\/} {\bf 732} L6 (\textit{Preprint} \eprint{arXiv:1101.4298
  [astro-ph.HE]})

\bibitem{Boyle:2007ft}
Boyle M, Barrow D~A, Kidder L~E, {Mrou\'e} A~H, Pfeiffer H~P, Scheel M~A, Cook
  G~B and Teukolsky S~A 2007 {\em Phys. Rev. D\/} {\bf 76} 124038
  (\textit{Preprint} \eprint{arXiv:0710.0158 [gr-qc]})

\bibitem{Blanchet:2013haa}
Blanchet L 2014 {\em Living Rev. Rel.\/} {\bf 17} 2 (\textit{Preprint}
  \eprint{1310.1528})

\bibitem{Damour2008}
Damour T, Nagar A, Hannam M, Husa S and Br{\"u}gmann B 2008 {\em Physical
  Review D\/} {\bf 78} 044039

\bibitem{buonanno:1999effective}
Buonanno A and Damour T 1999 {\em Phys. Rev.\/} {\bf D59} 084006
  (\textit{Preprint} \eprint{gr-qc/9811091})

\bibitem{Buonanno:2007pf}
Buonanno A, Pan Y, Baker J~G, Centrella J, Kelly B~J, McWilliams S~T and van
  Meter J~R 2007 {\em Phys. Rev. D\/} {\bf 76} 104049 (\textit{Preprint}
  \eprint{arXiv:0706.3732 [gr-qc]})

\bibitem{damour:2010effective}
Damour T and Nagar A 2010 {\em Phys. Rev.\/} {\bf D81} 084016
  (\textit{Preprint} \eprint{0911.5041})

\bibitem{Hinderer:2016eia}
Hinderer T {\em et~al.\/} 2016  (\textit{Preprint} \eprint{1602.00599})

\bibitem{Bernuzzi:2014owa}
Bernuzzi S, Nagar A, Dietrich T and Damour T 2015 {\em Phys. Rev. Lett.\/} {\bf
  114} 161103 (\textit{Preprint} \eprint{1412.4553})

\bibitem{Bernuzzi:2012ci}
Bernuzzi S, Nagar A, Thierfelder M and Brugmann B 2012 {\em Phys. Rev.\/} {\bf
  D86} 044030 (\textit{Preprint} \eprint{1205.3403})

\bibitem{Hotokezaka:2016bzh}
Hotokezaka K, Kyutoku K, Sekiguchi Y~i and Shibata M 2016  (\textit{Preprint}
  \eprint{1603.01286})

\bibitem{Haas:2016cop}
Haas R {\em et~al.\/} 2016  (\textit{Preprint} \eprint{1604.00782})

\bibitem{Szilagyi:2015rwa}
Szilágyi B, Blackman J, Buonanno A, Taracchini A, Pfeiffer H~P, Scheel M~A,
  Chu T, Kidder L~E and Pan Y 2015 {\em Phys. Rev. Lett.\/} {\bf 115} 031102
  (\textit{Preprint} \eprint{1502.04953})

\bibitem{Cactuscode:web}
{Cactus Computational Toolkit} \urlprefix\url{http://www.cactuscode.org/}

\bibitem{Goodale:2002a}
Goodale T, Allen G, Lanfermann G, Mass{\'o} J, Radke T, Seidel E and Shalf J
  2003 The {Cactus} framework and toolkit: Design and applications {\em Vector
  and Parallel Processing -- VECPAR'2002, 5th International Conference, Lecture
  Notes in Computer Science\/} (Berlin: Springer)
  \urlprefix\url{http://edoc.mpg.de/3341}

\bibitem{SVN:2016}
See the university of parma gravity group web page:
  http://www.fis.unipr.it/gravity/research/bns2015.html

\bibitem{CarpetCode:web}
{Carpet}: Adaptive Mesh Refinement for the {Cactus} Framework
  \urlprefix\url{http://www.carpetcode.org/}

\bibitem{Schnetter:2003rb}
Schnetter E, Hawley S~H and Hawke I 2004 {\em Class. Quantum Grav.\/} {\bf 21}
  1465--1488 (\textit{Preprint} \eprint{arXiv:gr-qc/0310042})

\bibitem{McLachlan:web}
{McLachlan}, a public {BSSN} code
  \urlprefix\url{http://www.cct.lsu.edu/~eschnett/McLachlan/}

\bibitem{Nakamura:1987zz}
Nakamura T, Oohara K and Kojima Y 1987 {\em Prog. Theor. Phys. Suppl.\/} {\bf
  90} 1--218

\bibitem{Shibata:1995we}
Shibata M and Nakamura T 1995 {\em Phys. Rev. D\/} {\bf 52} 5428--5444

\bibitem{Baumgarte:1998te}
Baumgarte T~W and Shapiro S~L 1999 {\em Phys. Rev. D\/} {\bf 59} 024007
  (\textit{Preprint} \eprint{arXiv:gr-qc/9810065})

\bibitem{Alcubierre:2000xu}
Alcubierre M, Br{\"u}gmann B, Dramlitsch T, Font J~A, Papadopoulos P, Seidel E,
  Stergioulas N and Takahashi R 2000 {\em Phys. Rev. D\/} {\bf 62} 044034
  (\textit{Preprint} \eprint{arXiv:gr-qc/0003071})

\bibitem{Alcubierre:2002kk}
Alcubierre M, Br{\"u}gmann B, Diener P, Koppitz M, Pollney D, Seidel E and
  Takahashi R 2003 {\em Phys. Rev. D\/} {\bf 67} 084023 (\textit{Preprint}
  \eprint{arXiv:gr-qc/0206072})

\bibitem{Harten:1987un}
{Harten} A, {Engquist} B, {Osher} S and {Chakravarthy} S~R 1987 {\em J. Comp.
  Phys.\/} {\bf 71} 231--303

\bibitem{Shu:1999ho}
Shu C~W 1999 High order {ENO} and {WENO} schemes for computational fluid
  dynamics {\em High order methods for computational physics\/} ed Barth T~J
  and Deconinck H~A (New York: Springer) pp 439--582

\bibitem{Harten:1983on}
Harten A, Lax P~D and van Leer B 1983 {\em SIAM review\/} {\bf 25} 35

\bibitem{Einfeldt:1988og}
Einfeldt B 1988 {\em SIAM J. Numer. Anal.\/} {\bf 25} 294--318

\bibitem{Runge:1895aa}
Runge C 1895 {\em Mathematische Annalen\/} {\bf 46} 167--178 ISSN 0025-5831
  \urlprefix\url{http://dx.doi.org/10.1007/BF01446807}

\bibitem{Kutta:1901aa}
Kutta W 1901 {\em Z. Math. Phys.\/} {\bf 46} 435--453

\bibitem{Kreiss:1973aa}
Kreiss H and Oliger J 1973 {\em {Methods for the Approximate Solution of Time
  Dependent Problems}\/} ({\em Global Atmospheric Research Programme (GARP):
  GARP Publication Series\/} vol~10) (GARP Publication)

\bibitem{Dietrich:2015pxa}
Dietrich T, Moldenhauer N, Johnson-McDaniel N~K, Bernuzzi S, Markakis C~M,
  Brügmann B and Tichy W 2015 {\em Phys. Rev.\/} {\bf D92} 124007
  (\textit{Preprint} \eprint{1507.07100})

\bibitem{Kyutoku:2014yba}
Kyutoku K, Shibata M and Taniguchi K 2014 {\em Phys. Rev.\/} {\bf D90} 064006
  (\textit{Preprint} \eprint{1405.6207})

\bibitem{Akmal:1998cf}
Akmal A, Pandharipande V and Ravenhall D 1998 {\em Phys. Rev. C\/} {\bf 58}
  1804--1828 (\textit{Preprint} \eprint{arXiv:nucl-th/9804027})

\bibitem{Douchin00}
{Douchin} F and {Haensel} P 2000 {\em Physics Letters B\/} {\bf 485} 107--114
  (\textit{Preprint} \eprint{arXiv:astro-ph/0006135})

\bibitem{Douchin01}
{Douchin} F and {Haensel} P 2001 {\em Astron. Astrophys.\/} {\bf 380} 151--167
  (\textit{Preprint} \eprint{arXiv:astro-ph/0111092})

\bibitem{Lackey:2005tk}
Lackey B~D, Nayyar M and Owen B~J 2006 {\em Phys. Rev.\/} {\bf D73} 024021
  (\textit{Preprint} \eprint{astro-ph/0507312})

\bibitem{Muller:1995ji}
Muller H and Serot B~D 1995 {\em Phys. Rev.\/} {\bf C52} 2072--2091
  (\textit{Preprint} \eprint{nucl-th/9505013})

\bibitem{Demorest:2010bx}
Demorest P, Pennucci T, Ransom S, Roberts M and Hessels J 2010 {\em Nature\/}
  {\bf 467} 1081--1083 (\textit{Preprint} \eprint{1010.5788})

\bibitem{Stergioulas95}
Stergioulas N and Friedman J~L 1995 {\em Astrophys. J.\/} {\bf 444} 306

\bibitem{Read:2009constraints}
Read J~S, Lackey B~D, Owen B~J and Friedman J~L 2009 {\em Physical Review D\/}
  {\bf 79} 124032

\bibitem{bauswein:2010testing}
Bauswein A, Janka H~T and Oechslin R 2010 {\em Physical Review D\/} {\bf 82}
  084043

\bibitem{Newman:1961qr}
Newman E and Penrose R 1962 {\em J. Math. Phys.\/} {\bf 3} 566--578

\bibitem{Baker:2001sf}
Baker J~G, Campanelli M and Lousto C~O 2002 {\em Phys. Rev. D\/} {\bf 65}
  044001 (\textit{Preprint} \eprint{arXiv:gr-qc/0104063})

\bibitem{Thorne:1980ru}
Thorne K~S 1980 {\em Rev. Mod. Phys.\/} {\bf 52} 299--339

\bibitem{Reisswig:2011notes}
Reisswig C and Pollney D 2011 {\em Classical and Quantum Gravity\/} {\bf 28}
  195015

\bibitem{Berti:2007inspiral}
Berti E, Cardoso V, Gonzalez J~A, Sperhake U, Hannam M, Husa S and Br{\"u}gmann
  B 2007 {\em Physical Review D\/} {\bf 76} 064034

\bibitem{Nakano:2015perturbative}
Nakano H, Healy J, Lousto C~O and Zlochower Y 2015 {\em Physical Review D\/}
  {\bf 91} 104022

\bibitem{Brugmann:2008calibration}
Br{\"u}gmann B, Gonz{\'a}lez J~A, Hannam M, Husa S, Sperhake U and Tichy W 2008
  {\em Physical Review D\/} {\bf 77} 024027

\bibitem{Boyle:2009vi}
Boyle M and Mroue A~H 2009 {\em Phys. Rev.\/} {\bf D80} 124045
  (\textit{Preprint} \eprint{0905.3177})

\bibitem{Lousto:2010qx}
Lousto C~O, Nakano H, Zlochower Y and Campanelli M 2010 {\em Phys. Rev.\/} {\bf
  D82} 104057 (\textit{Preprint} \eprint{1008.4360})

\bibitem{Bishop:1998uk}
Bishop N~T, Gomez R, Lehner L, Szilagyi B, Winicour J and Isaacson R~A 1998
  (\textit{Preprint} \eprint{arXiv:gr-qc/9801070})

\bibitem{Babiuc:2010ze}
Babiuc M, Szilagyi B, Winicour J and Zlochower Y 2011 {\em Phys. Rev. D\/} {\bf
  84} 044057 (\textit{Preprint} \eprint{arXiv:1011.4223 [gr-qc]})

\bibitem{Lousto:2013oza}
Lousto C~O and Zlochower Y 2013 {\em Phys. Rev.\/} {\bf D88} 024001
  (\textit{Preprint} \eprint{1304.3937})

\bibitem{Hinder:2013oqa}
Hinder I {\em et~al.\/} 2014 {\em Class. Quant. Grav.\/} {\bf 31} 025012
  (\textit{Preprint} \eprint{1307.5307})

\bibitem{Chu:2015kft}
Chu T, Fong H, Kumar P, Pfeiffer H~P, Boyle M, Hemberger D~A, Kidder L~E,
  Scheel M~A and Szilagyi B 2015  (\textit{Preprint} \eprint{1512.06800})

\bibitem{Kinnersley:1969zza}
Kinnersley W 1969 {\em J. Math. Phys.\/} {\bf 10} 1195--1203

\bibitem{Nakano:2015rda}
Nakano H 2015 {\em Class. Quant. Grav.\/} {\bf 32} 177002 (\textit{Preprint}
  \eprint{1501.02890})

\bibitem{Tanay:2016zog}
Tanay S, Haney M and Gopakumar A 2016 {\em Phys. Rev.\/} {\bf D93} 064031
  (\textit{Preprint} \eprint{1602.03081})

\bibitem{Radice:2016gym}
Radice D, Bernuzzi S and Ott C~D 2016  (\textit{Preprint} \eprint{1603.05726})

\bibitem{Lindblom:2008cm}
Lindblom L, Owen B~J and Brown D~A 2008 {\em Phys. Rev.\/} {\bf D78} 124020
  (\textit{Preprint} \eprint{0809.3844})

\bibitem{Boyle:2008ge}
Boyle M, Buonanno A, Kidder L~E, Mroue A~H, Pan Y, Pfeiffer H~P and Scheel M~A
  2008 {\em Phys. Rev.\/} {\bf D78} 104020 (\textit{Preprint}
  \eprint{0804.4184})

\bibitem{Baiotti:2010xh}
Baiotti L, Damour T, Giacomazzo B, Nagar A and Rezzolla L 2010 {\em Phys. Rev.
  Lett.\/} {\bf 105} 261101 (\textit{Preprint} \eprint{arXiv:1009.0521
  [gr-qc]})

\bibitem{Suh:2016ctx}
Suh I, Mathews G~J, Haywood J~R and Lan N~Q 2016  (\textit{Preprint}
  \eprint{1601.01460})

\bibitem{Damour:2011fu}
Damour T, Nagar A, Pollney D and Reisswig C 2012 {\em Phys. Rev. Lett.\/} {\bf
  108} 131101 (\textit{Preprint} \eprint{1110.2938})

\bibitem{Alic:2011gg}
Alic D, Bona-Casas C, Bona C, Rezzolla L and Palenzuela C 2012 {\em Phys.
  Rev.\/} {\bf D85} 064040 (\textit{Preprint} \eprint{1106.2254})

\bibitem{Weyhausen:2011cg}
Weyhausen A, Bernuzzi S and Hilditch D 2012 {\em Phys. Rev.\/} {\bf D85} 024038
  (\textit{Preprint} \eprint{1107.5539})

\bibitem{Radice:2012cu}
Radice D and Rezzolla L 2012 {\em Astron. Astrophys.\/} {\bf 547} A26
  (\textit{Preprint} \eprint{1206.6502})

\bibitem{DePietri:2014mea}
De~Pietri R, Feo A, Franci L and Löffler F 2014 {\em Phys. Rev.\/} {\bf D90}
  024034 (\textit{Preprint} \eprint{1403.8066})

\bibitem{Loffler:2014jma}
Löffler F, De~Pietri R, Feo A, Maione F and Franci L 2015 {\em Phys. Rev.\/}
  {\bf D91} 064057 (\textit{Preprint} \eprint{1411.1963})

\bibitem{Rezzolla:2016nxn}
Rezzolla L and Takami K 2016  (\textit{Preprint} \eprint{1604.00246})

\bibitem{bernuzzi:2015modeling}
Bernuzzi S, Dietrich T and Nagar A 2015 {\em Phys. Rev. Lett.\/} {\bf 115}
  091101 (\textit{Preprint} \eprint{1504.01764})

\bibitem{Stergioulas:2011gd}
Stergioulas N, Bauswein A, Zagkouris K and Janka H~T 2011 {\em Mon. Not. Roy.
  Astron. Soc.\/} {\bf 418} 427 (\textit{Preprint} \eprint{1105.0368})

\end{thebibliography}
\end{document}